\documentstyle[prd,preprint,aps]{revtex}

\begin{document}

% Here comes the title page ***************************************************
%\draft
\title{
                   Solar Model Uncertainties, MSW Analysis, and\\
		   Future Solar Neutrino Experiments
}
\author{
                   Naoya Hata and Paul Langacker
}
\address{                                                         %**revtex**
                   Department of Physics,                         %**revtex**
                   University of Pennsylvania,   \\               %**revtex**
                   Philadelphia, Pennsylvania 19104               %**revtex**
}                                                                 %**revtex**
\date{
%                   \today, Draft Version 2.5
                   November 2, 1993, UPR-0592T
}
\maketitle
%
%
%% Here comes the abstract ***************************************************
%
%
\begin{abstract}

%\begin{center}\parbox[h]{0.50\textwidth}{                         %**2column**

Various theoretical uncertainties in the standard solar model
and in the Mikheyev-Smirnov-Wolfenstein (MSW) analysis are discussed.
It is shown that two methods of estimating the solar neutrino flux
uncertainties are equivalent:  (a) a simple parametrization of the
uncertainties using the core temperature and the nuclear production
cross sections; (b) the Monte Carlo method of Bahcall and Ulrich.
In the MSW analysis, we emphasize proper treatments of correlation
of theoretical uncertainties between flux components and between
different detectors, the Earth effect, and multiple solutions in
a combined $\chi^2$ procedure.  The MSW solutions for various
standard and nonstandard solar models are also shown.  The MSW
predictions of the global solutions for the future solar neutrino
experiments are given, emphasizing the measurement of the energy
spectrum and the day-night effect in Sudbury Neutrino Observatory
and Super-Kamiokande to distinguish the two solutions.

%}\end{center}                                                     %**2column**

\end{abstract}
\pacs{PACS numbers: 96.60.Kx, 12.15.Fp, 14.60.Gh}                %**revtex**

\newpage
%
% Here comes text *************************************************************
%
%\twocolumn                                                       %**2column**

\section{%
             Introduction %
}
\label{sec_introduction}

The solar neutrino experiments of Homestake (chlorine) \cite{Homestake,%
Homestake-update}, Kamiokande \cite{Kamiokande,Kamiokande-update}, and the
gallium experiments of SAGE \cite{SAGE,SAGE-update}
and GALLEX \cite{GALLEX} show  deficits of the neutrino flux from the
Sun when compared to the standard solar model (SSM) predictions
\cite{Bahcall-Pinsonneault,Turck-Chieze-Lopes} as summarized in
Table~\ref{tab_exps}.  Numerous theoretical proposals have been made to
resolve the discrepancy between theory and experiment.  Astrophysical
solutions in general are, however, strongly disfavored by the data.  As long
as astrophysical processes cannot significantly distort the neutrino energy
spectrum \cite{Bahcall-spectrum} the lower observed Homestake rate relative
to the Kamiokande rate excludes essentially all astrophysical explanations
\cite{BHL,BKL,BHKL,Bahcall-Bethe}.  Even with the Homestake experimental error
tripled, the combined observations are in contradiction with the explicitly
constructed nonstandard solar models \cite{BHL}.  On the other hand, among
many particle physics solutions, the Mikheyev-Smirnov-Wolfenstein (MSW)
mechanism \cite{MSW} gives an excellent description of the data, and is taken
as a strong hint of neutrino mass and mixings \cite{BHKL,HL}.  (See also
\cite{Bahcall-Haxton,Shi-Schramm,Gelb-Kwong-Rosen,Krastev-Petcov,%
Krauss-Gates-White}.)

When constraining the parameter space from the experimental data in the
MSW analysis, it is necessary to include relevant theoretical uncertainties
properly.   In the SSM, the theoretical uncertainty in the initial
$^8$B flux quoted by Bahcall and Pinsonneault \cite{Bahcall-Pinsonneault} is
14\%, and is comparable to the experimental uncertainties of Homestake
(10\%) and Kamiokande (14\%).  In the Turck-Chi\`eze--Lopes SSM
\cite{Turck-Chieze-Lopes}, the $^8$B flux uncertainty is 25\% and dominates
the experimental uncertainties.  The omission of the theoretical
uncertainties underestimates the uncertainty of the MSW parameter
space constrained from the experiments.

Equally important, but often ignored, are the correlations among the
theoretical uncertainties.  Especially, a correct treatment of the $^8$B flux
uncertainty is significant since it is the largest among the theoretical
uncertainties and also strongly correlated from experiment to experiment.
When one considers the combined fit, for example, it is not legitimate to
allow a smaller $^8$B flux for Homestake and a larger $^8$B flux for
Kamiokande.  If the correlations were ignored in the MSW two-flavor
oscillation analysis, one obtains a larger parameter space in the
large-angle solution and even finds a third allowed region around
$\Delta m^2 \sim 10^{-7}\, \mbox{eV}\, ^2$ and $\sin^22\theta \sim 0.7$ at
90\% C.L.  Moreover, the uncertainties are also correlated among different
flux components.  For instance, if the opacity were lower than the standard
value (or equivalently the core temperature were lower), both the $^7$Be
and $^8$B fluxes would be reduced.  Since the MSW mechanism often
affects each neutrino flux component differently according to the
neutrino energy, the flux uncertainties and their correlations affect
nontrivially  the allowed parameter space of the combined observations.

Two methods have been proposed to incorporate the flux uncertainties in the
MSW analysis. The first method utilizes the Monte Carlo SSMs constructed with
randomly chosen input parameters distributed around the mean  values
\cite{Bahcall-Ulrich}.  Those 1000 SSMs were incorporated in the MSW analysis
of Bahcall and Haxton \cite{Bahcall-Haxton}, and, recently, of Shi, Schramm,
and Bahcall \cite{Shi-Schramm}.  Krauss, Gates, and White calculated the flux
uncertainties from their own 100,000 Monte Carlo SSMs
\cite{Krauss-Gates-White}.
\footnote{
The correlations among flux uncertainties are ignored in the analysis in
Ref.~\cite{Krauss-Gates-White}.  Also the uncertainties used in
Ref.~\cite{Krauss-Gates-White} are larger than the estimation of Bahcall
and Pinsonneault.
}
This Monte Carlo method provides a rigorous estimation of the SSM
uncertainties as long as their correlations are properly taken into account
for different flux components and different experiments.

The second method, which we have used in our previous analysis
\cite{BHKL,HL},  parametrizes the SSM flux uncertainties with the central
temperature and nuclear reaction cross sections, distinguishing the flux
uncertainties due to purely astrophysical effects, such as the uncertainties
from the heavy element abundance and other uncertainties in the opacity, from
those due to
the nuclear cross sections.  The astrophysical uncertainties are parametrized
by the uncertainty of the central temperature $\Delta T_C$, which is chosen to
reproduce the uncertainties derived from the Monte Carlo
estimation.  This parametrization method has several advantages over the
Monte Carlo method.  The physical meaning of the uncertainties is clear, and
it can be generalized to the nonstandard solar models that are in most cases
parametrized by the lower $T_C$ \cite{Bahcall-Ulrich,BHL}.  Also one can
easily
update the calculations by just changing the values of $\Delta T_C$ and
nuclear cross section errors when a new solar model is introduced.  It
required only trivial changes to update our calculations from the
Bahcall-Ulrich model to the Bahcall-Pinsonneault model, while the Monte Carlo
SSMs are not yet available for the latter.

This parametrization of the uncertainties, however, was questioned on the
grounds that such a simplification can lead to errors in describing the
nonlinear relations among the neutrino fluxes, which are the output of
solving the coupled partial differential equations of stellar structure with
nontrivial matching conditions \cite{Bahcall-Texas,Bahcall-Bethe}.  It was
argued that a Monte Carlo study is necessary to estimate the uncertainties
reliably.  Later we will show numerically that, in fact, the uncertainties
obtained by he parametrization method are essentially identical to the
Monte Carlo results.

In this paper we discuss various technical but important issues concerning
the theoretical uncertainties in the solar neutrino data analysis.
In Section~\ref{sec_uncertainty}, we show that the parametrization method
reproduces the SSM uncertainties obtained by the Monte Carlo method.  The
comparison is made for both the uncertainties of flux components and their
correlations.  We also compare the two methods for the uncertainties and
their correlations of the predicted rates for different solar neutrino
detectors.

Once the equivalence of the two methods is established
for the SSM, we compare them in the MSW analysis in
Section~\ref{sec_MSW}.  The allowed regions calculated by both
methods are shown.  The effect of the theoretical uncertainties
and their correlations are displayed for the global MSW analysis.
Other theoretical issues are also considered in Section~\ref{sec_MSW}.
Analytic approximations for the MSW transitions by Petcov, Parke,
and Pizzochero are compared and the associated uncertainties are discussed.
Another issue involves the estimate of confidence level (C.L.) in the
presence of multiple fit solutions.  We emphasize that there are several
possible definitions of the C.L. contours, leading to slightly
different allowed regions, and give a statistical definition of
the C.L. contours of the most conservative (and we believe the best)
prescription.  We also compare the allowed MSW parameter space for the
SSM with increased uncertainties and the SSMs of different authors.

We conclude Section~\ref{sec_MSW} by presenting the current results of
our MSW fits for transitions into both ordinary ($\nu_\mu, \nu_\tau$)
and sterile neutrinos, incorporating the theoretical issues and
uncertainties discussed above.  These also include the regeneration in
the Earth (Earth effect), which is important for some regions of the
MSW parameters for both time-averaged data and day-night asymmetries
\cite{Earth-effect,HL}.  We find that the data is fit extremely well by
the MSW effect for transitions into ordinary neutrinos for
$\Delta m^2 \sim 6 \times 10^{-6} \, \mbox{eV}\, ^2$ and $\sin^22\theta
\sim 7 \times 10^{-3}$ (the nonadiabatic) solution, although there is a
second (large-angle) solution with $\Delta m^2 \sim 9 \times 10^{-6} \,
\mbox{eV}\, ^2$ and $\sin^22\theta \sim 0.6$ which is marginally
allowed at 90\% C.L.  There is also a poorer but acceptable fit for
transitions into sterile neutrinos in the nonadiabatic region.

The MSW effect can be also considered in the nonstandard solar models, and
the combined fits to explicitly
constructed nonstandard solar models are shown in Section~\ref{sec_nonssms}.
The results of the MSW fit using the core temperature and the $^8$B flux
each as a free parameter are displayed. In Section~\ref{sec_future},
the prospects for the future solar neutrino experiments are considered.
We discuss in detail the predictions for the energy spectrum measurement
and the Earth effect in Sudbury Neutrino Observatory (SNO) and
Super-Kamiokande to distinguish the two solutions obtained from the global
analysis:  a spectrum distortion is predicted for the nonadiabatic
solution, while a characteristic day-night effect is expected for
the large-angle solution.

\section{%
               Comparison of the Parametrized Uncertainties
               and the Monte Carlo Results                    %
}
\label{sec_uncertainty}

\subsection{%
               The Flux Uncertainties  %
}

In our parametrization method \cite{BHKL,HL}, the basic assumption is
that the SSM flux
uncertainties are expressed by a few parameters that have physical meanings,
and that those uncertainties are approximated by gaussian distributions.
First we studied the distributions of the 1000 Monte Carlo SSM fluxes
calculated by Bahcall and Ulrich with randomly chosen input parameters
\cite{Bahcall-Ulrich}, and confirmed that they indeed have gaussian
distributions.  In Fig.~\ref{fig_gaussian_fluxes}, the histograms are the
distribution of the Monte Carlo fluxes of the $pp$, $^7$Be, and $^8$B
neutrinos; the solid lines are our fit with the gaussian form.  The rate
predictions for solar neutrino detectors were also studied and are displayed
in Fig.~\ref{fig_gaussian_exps}, which are also fit with the gaussian form.
In both cases, the fits are excellent.

Secondly, we utilize the observation that the neutrino fluxes are well
described by a power law in  the central temperature \cite{T_c-error},
\begin{equation}
	\phi(pp)        \sim T_C^{-1.2}, \;
	\phi(\mbox{Be}) \sim T_C^{8},	\; \mbox{ and } \;
	\phi(\mbox{B})  \sim T_C^{18},
\label{eq_power-law}
\end{equation}
where our units are such that $T_C = 1 = 15.67 \times 10^6 \;\mbox{K}$ for
the central value of the SSM.
We express the astrophysical uncertainties of the major fluxes ($pp$, $^7$Be,
and $^8$B) with the uncertainty of the central temperature times the exponent
in the power law.  We identify the main sources of the astrophysical
uncertainties as the heavy element abundance and other uncertainties
in the opacity.
The other uncertainties, independent of such astrophysical effects, are from
the nuclear reaction cross sections, especially for the cross
sections of $p + ^7\!\mbox{Be}$ and $^3\mbox{He} + ^4\!\mbox{He}$.  Those
cross sections are expressed by S-factors $S_{17}$ and  $S_{34}$,
respectively.
\footnote{
While the $p+^7\!\mbox{Be}$ cross section has little effect on solar
conditions other than the $^8$B flux, other cross sections such as $p+p$
($S_{11}$) and $^3\mbox{He}+^3\!\mbox{He}$ ($S_{33}$) directly affect
the energy generation and cannot be separated from the astrophysical
uncertainties.  In particular, the $p+p$ reaction is the main source of the
energy production, and a change of $S_{11}$ leads to a change
in $T_C$.  We therefore consider that the effects of the uncertainties in
$S_{11}$ and $S_{33}$ are included in $\Delta T_C$.
}
Our choice of the relevant cross sections is based on the fact that
the $^8$B flux is
directly proportional to  $S_{17}$, and  the $^7$Be and $^8$B fluxes are
proportional to  $S_{34}$ \cite{Bahcall-logderi}.  The flux uncertainties
and their correlations are expressed by those three parameters
\footnote{
We could extend this parametrization method by taking into account all nine
SSM input parameters using the partial derivatives of the neutrino fluxes
obtained from the Monte Carlo SSMs \cite{Bahcall-logderi}.  This should
completely reproduce the Monte Carlo results.  We will show, however,
that our minimal choice of the parameters is sufficient to describe
the SSM uncertainties.
}
($\Delta T_C$, $\Delta S_{17}/S_{17}$, and $\Delta S_{34}/S_{34}$).
Their contribution to each flux is displayed in Table~\ref{tab_parameters}.
The magnitude of the uncertainties of the major fluxes
\footnote{
We similarly parametrized the minor fluxes [$^{13}$N, $^{15}$O, $^{17}$F,
$p + e + p$ ($pep$), and $^3\mbox{He} + p$ ($hep$)] using $T_C$ exponents
$n = 22, 28.5, 28.8, 2.8 \mbox{ and } 4.5$, respectively, and reproduce the
amplitude of the uncertainties.  However, the $T_C$ exponents for these fluxes
except $hep$ are not obtained in Ref.~\cite{Bahcall-Ulrich}, and we do not
reproduce the correlations of the $pep$ and $hep$ neutrinos with others
properly.  In the MSW calculations, the
effect of the correlations among the minor fluxes is completely negligible.
}
are the quadrature sum of those uncertainties:
\begin{equation}
    \frac{\Delta \phi^i}{\phi^i}
       = \left[  ( n_i \Delta T_C )^2 +
                   \sum_{k=34,17} ( s_k^i )^2
         \right]^{1/2},
\end{equation}
where $i= pp, ^7\!\mbox{Be}, ^8\!\mbox{B}$. $n_i$ are the temperature
exponents
($\phi^i \sim T_C^{n_i}$), and $s_k^i$ are the fractional  uncertainty in
$\phi^i$ from $\Delta S_{17}/S_{17}$ and $\Delta S_{34}/S_{34}$
listed in Table~\ref{tab_parameters} for both the
Bahcall-Ulrich and Bahcall-Pinsonneault model.  $\Delta T_C$ is not explicitly
given in the SSM calculation.  We determine $\Delta T_C$ so that the
uncertainties defined above correctly reproduce the Bahcall-Ulrich or
Bahcall-Pinsonneault flux uncertainties; the resulting  $\Delta T_C$ is
consistent with the $T_C$ distribution of the Monte Carlo SSMs displayed in
Ref.~\cite{T_c-error}.  We obtained $\Delta T_C = 0.0057$ for the
Bahcall-Ulrich model, and $\Delta T_C = 0.0060$ for the Bahcall-Pinsonneault
model.  (The Monte Carlo SSMs of the Bahcall-Pinsonneault model are not yet
available.)

The correlation matrix of the flux uncertainties is given by
\begin{equation}
\label{eqn_cij_parm}
      C_{ij} =  \left(
                     n_i n_j (\Delta T_C)^2 +
                     \sum_{k=34,17} s^i_k s^j_k
                 \right) / \left(
                      \frac{ \Delta \phi^i }{ \phi^i }
                      \frac{ \Delta \phi^j }{ \phi^j }
                 \right) ,
\end{equation}
where $i, j = pp, ^7\!\mbox{Be}, ^8\!\mbox{B}$.

The result of those uncertainties are compared to the Bahcall-Ulrich Monte
Carlo SSMs, whose flux  uncertainties are calculated by
\begin{equation}
    \frac{\Delta \phi^i}{\bar{ \phi }^i}
       = \left[  \frac{1}{N}
                 \sum_{m=1}^N
                   \left( \frac{ \phi^i_m }{ \bar{ \phi }^i} - 1 \right)^2
         \right]^{1/2},
\end{equation}
where $\phi^i_m $ are the $i$th fluxes of the Monte Carlo SSMs
($m = 1, \cdots , N=1000$); $\bar{ \phi^i } [ = (\sum_{m=1}^N \phi^i_m ) / N]$
is the mean value of the Monte Carlo fluxes.  The correlation matrix is
obtained by
\begin{equation}
\label{eqn_cij_mc}
      C_{ij} =   \frac{ 1 }{ N }
                 \sum_{m=1}^N
                    \left( \frac{ \phi^i_m }{ \bar{ \phi^i } } - 1 \right)
                    \left( \frac{ \phi^j_m }{ \bar{ \phi^j } } - 1 \right)
                  /
                 \left(
                      \frac{ \Delta \phi^i }{ \bar{ \phi^i } }
                      \frac{ \Delta \phi^j }{ \bar{ \phi^j } }
                 \right) .
\end{equation}

The results of the two calculations are compared in Table~\ref{tab_magnitudes}
for the magnitudes of the uncertainties and Table~\ref{tab_correlations} for
the correlations.  The agreement of the magnitudes are excellent.  The
correlation matrices are also in good agreement especially for the $pp$-Be
element.  In Fig.~\ref{fig_correlations}, we display the distribution of
Monte Carlo fluxes and our parametrization in $\phi(\mbox{Be}) -
\phi(\mbox{B})$ plane; the agreement of the two methods are remarkable.

\subsection{%
	      The Rate Uncertainties %
}

The comparison of the two methods is also carried out for the predictions
for different solar neutrino detectors.  The fraction of contribution
$f^d_i$ of the $i$th flux component to the $d$th detector ($d=$Kamiokande, Cl,
Ga) are listed in Table~\ref{tab_fractions}.  The SSM uncertainty of the rate
$R^d$ (in units of the central values of the SSM) for each detector is a
quadratic sum of the flux uncertainties and the detector cross section
uncertainties $ \Delta \sigma^d / \sigma^d $, which is 0.033 and 0.04 for
the Cl and Ga detector, respectively \cite{Bahcall-Ulrich};
\begin{equation}
	 \Delta R^d  =
            \left[
                  \sum_{i=\mbox{{\scriptsize fluxes}}}
                      \left(
                         f^d_i \, \frac{ \Delta \phi^i }{ \phi^i }
                      \right)^2
                   +  \left(
                         \frac{ \Delta \sigma^d }{ \sigma^d }
                      \right)^2
            \right]^{1/2}   .
\end{equation}
Those uncertainties are correlated by the $\Delta T_C$ and $s^i_k$
through the fluxes; the error matrix is
\begin{equation}
        V_{cd}   =   \sum_{i,j=\mbox{{\scriptsize fluxes}}} f^c_i \, f^d_j \,
                   [ \,
                      n^i \, n^j ( \Delta T_C )^2
                    + \sum_{k=34,17} s^i_k \, s^j_k \,
                   ],
\end{equation}
where $c, d = \mbox{Kamiokande, Cl, Ga}$.  Here (and below) the correlation
matrix $D_{cd}$ is related to the error matrix by
\begin{equation}
	D_{cd}   =   {V_{cd} \over \Delta R^c \Delta R^d}  .
\end{equation}
Among the correlations, the $^8$B flux uncertainty is most
significant because of its large amplitude and strong correlation between the
experiments, especially between the Kamiokande and Cl rate.

For the Monte Carlo method,  the rate for the $m$th Monte Carlo SSM for the
$d$th detector is given by
\begin{equation}
        R^d_m       =  \sum_{i=\mbox{\scriptsize fluxes}}
	                    f^d_i \, \frac{ \phi^i_m }{ \bar{\phi^i} }.
\end{equation}
The rate uncertainties for the $d$th detector is
\begin{equation}
	\Delta R^d  =  \left[ {1 \over N} \sum_{m=1}^N ( R^d_m - 1 )^2
                       \right]^{1/2} .
\end{equation}
The error matrix element between the $c$th and $d$th detectors is
\begin{equation}
	V_{cd} = \frac{ 1 }{ N }
                 \sum_{m=1}^N
	           ( R^c_m  -  1 ) (  R^d_m - 1) .
\end{equation}

We compare the uncertainties and their correlations for the rates in
Tables~\ref{tab_exp_magnitudes} (magnitudes) and \ref{tab_exp_correlations}
(correlations).  The parametrization method reproduces the Monte Carlo
results with a remarkable accuracy.

\section{%
                The Theoretical Uncertainties in The MSW Analysis        %
}
\label{sec_MSW}
\subsection{%
                The SSM Uncertainties  %
}

We incorporate the SSM flux uncertainties described above in the MSW analysis
using a $\chi^2$ method.  The MSW calculations of the theoretical rate
predictions for each experiment are described in Refs.~\cite{BHKL,HL}. (See
also Refs.~\cite{Kennedy,Parke,Pizzochero,Petcov,Haxton,Kuo-Pantaleone}.)
The MSW rate for the $d$th detector ($d = \mbox{Kamiokande, Cl, Ga}$) is
\begin{equation}
	R^{d}_{\mbox{{\scriptsize MSW}}} (\sin^22\theta,\Delta m^2)
	         = \sum_{i=\mbox{{\scriptsize fluxes}}}
                        f^d_i \, P_i^d(\sin^22\theta,\Delta m^2),
\end{equation}
where $P_i^d$ is the MSW survival probability of the fluxes
($i$ represents nine flux components, $pp$, $^7$Be(I), $^7$Be(II), $^8$B,
$^{13}$N, $^{15}$O, $^{17}$F, $pep$ and $hep$) after integrating over the
neutrino production site and the neutrino energy including the detector
cross sections;  for Kamiokande the detector resolution and efficiency are
also included when integrating over the recoil electron energy.  The
formula is also valid
with the time-averaged Earth effect (but not with the Kamiokande
II day-night data with six time bins).  We calculate a $\chi^2$ value
for each point in the
$\sin^22\theta - \Delta m^2$
parameter space;
\begin{equation}
	\chi^2 (\Delta m^2, \sin^22\theta) =
                 \sum_{c,d = \mbox{{\scriptsize Kam, Cl, Ga}}}
	                        ( R^c_{\mbox{{\scriptsize expt}}}
                                - R^c_{\mbox{{\scriptsize MSW}}} )\;
				(V^{-1})_{cd}                     \;
                                ( R^d_{\mbox{{\scriptsize expt}}}
                                - R^d_{\mbox{{\scriptsize MSW}}} ),
\end{equation}
where $R_{\mbox{{\scriptsize expt}}}$ are the experimental values listed in
Table~\ref{tab_exps}.   $V$ is the $3 \times 3$ error matrix and its diagonal
elements are the quadratic sum of the experimental uncertainties, the detector
cross section uncertainties, and the SSM flux uncertainties:
\begin{equation}
\label{eqn_error-one}
	V_{dd} =  (\Delta R^d_{\mbox{{\scriptsize expt}}})^2 +
                \left(
                      \frac{ \Delta \sigma^d }{ \sigma^d }
                       R^d_{\mbox{{\scriptsize MSW}}}
                \right)^2
               + \sum_{i=\mbox{{\scriptsize fluxes}}}
                    \left(
                         P_i \,  f^d_i \, \frac{ \Delta \phi^i }{ \phi^i }
                    \right)^2
\end{equation}
The off-diagonal elements describe the correlations of the flux
uncertainties described by $\Delta T_C$ and $s^i_k$;
\begin{equation}
\label{eqn_error-two}
        V_{cd} = \sum_{i,j=\mbox{{\scriptsize fluxes}}}
                       P_i \, P_j \, f^c_i \, f^d_j \,
                  [ \,
                      n^i \, n^j ( \Delta T_C )^2
                    + \sum_{k=34,17} s^i_k \,  s^j_k \,
                   ].
\end{equation}

For the Monte Carlo SSMs, the $\chi^2$ is defined by a Monte Carlo average of
the probability function:
\begin{equation}
	\exp ( -\chi^2 / 2 )  =
                      \frac{1}{N} \sum_{m=1}^N \exp ( -\chi^2_m / 2 )
\end{equation}
where $\chi^2_m$ is the $\chi^2$ value calculated for the $m$th Monte Carlo
SSM ($m = 1, \cdots, N=1000$);
\begin{equation}
        \chi^2_m  =
	           \sum_{d=\mbox{{\scriptsize Kam,Cl,Ga}}}
		      \frac{ ( R^d_{\mbox{{\scriptsize expt}}}
                             - R^d_{m, \mbox{{\scriptsize MSW}}} )^2 }
                           { ( \Delta R^d_{\mbox{{\scriptsize expt}}} )^2 +
                             ( R^d_{m, \mbox{{\scriptsize MSW}}} \,
                               \Delta \sigma^d / \sigma^d
                              )^2 },
\end{equation}
with the MSW predicted rate for the $m$th Monte Carlo:
\begin{equation}
	R^d_{m, \mbox{{\scriptsize MSW}}
                     } = \sum_{i=\mbox{{\scriptsize fluxes}}}
                         P_i \, f^d_i \, \frac{ \phi^i_m }{ \bar{ \phi^i } }.
\end{equation}

We compare
\footnote{%
The 95\% C.L. allowed regions in Figs.~\ref{fig_MSW-noearth},
\ref{fig_MSW-earth}, \ref{fig_MSW-uncertainties},
\ref{fig_TC-SSM}, \ref{fig_larger-errors},
\ref{fig_sterile}, \ref{fig_MSW-nonssms}, \ref{fig_MSW-nonssms2},
\ref{fig_MSW-Tcfree}, \ref{fig_MSW-B8free},
\ref{fig_Ga-prediction}, and \ref{fig_future-detectors} are defined by
$\chi^2 (\sin^22\theta, \Delta m^2) \leq \chi^2_{\mbox{\scriptsize min}}
+ \Delta \chi^2$ with $\Delta \chi^2 = 6.0$ for both combined fits and
individual experiment fits, which corresponds to Gaussian errors in
two parameters.  Improved definitions for $\Delta \chi^2$ in
the combined fits  will be discussed later.  For the individual experiments
we use $\Delta \chi^2 = 6.0$ (2 d.f.) instead of 3.9 (1 d.f.), since it
is a more conservative estimate of the uncertainties and also easier to
compare with the combined fits.  (The latter would correspond to mapping
a one-parameter confidence region onto a band in a two-parameter space.)
}
the Monte Carlo result and the parametrized method for both
the Bahcall-Ulrich model and the Bahcall-Pinsonneault model in
Fig.~\ref{fig_MSW-noearth} when the Earth effect is ignored.  The agreement
between the Monte Carlo SSMs and the Bahcall-Ulrich model with our
parametrization is excellent.

It is important to include the possibility of $\nu_e$ regeneration in
the Earth \cite{Earth-effect,HL}.  This affects the time-averaged rates
and in addition, Kamiokande has searched for a day-night asymmetry by
binning their data with respect to the angle between the nadir and the
Sun.  No asymmetry was observed.

One can easily generalize Eqns~\ref{eqn_error-one} and \ref{eqn_error-two}
to include the Kamiokande day-night data point \cite{Kam-daynight}
by expanding the error matrix to 9 $\times$ 9, representing the time-averaged
rates of  Homestake, gallium, Kamiokande III, and the six Kamiokande
II day-night data points \cite{HL}.  We have scaled the normalized
Kamiokande II day-night data taken from Ref.~\cite{Kam-daynight} to
the quoted Kamiokande II average
value.  Also we have added to each of the six bins the overall systematic
uncertainty (15\%) from the energy calibration, the angular resolution, and
the event selection, which are factored out in the quoted normalized data.
(We checked consistency by combining the six bins and reproducing the
quoted average Kamiokande II rate.)  The systematic uncertainty as well
as the SSM flux uncertainties are properly correlated among the six bins
and the other uncertainties.

Fig.~\ref{fig_MSW-earth} shows the MSW allowed regions
for the three SSMs when the Earth effect is included. Again we conclude that
the two methods yield essentially the same results, although the obtained
large-angle region is slightly smaller in the parametrization method.

We demonstrate the effect of the theoretical uncertainties by comparing the
allowed MSW regions calculated without the flux and detector cross
section uncertainties.  Fig.~\ref{fig_MSW-uncertainties}(a) is the result
without the theoretical uncertainties using the Bahcall-Pinsonneault SSM.
When compared to Fig.~\ref{fig_MSW-earth}(c), the nonadiabatic allowed
region is noticeably smaller; there is no large-angle solution at 95\% C.L.

The omission of the correlation of uncertainties among the experiments
can lead to a overestimation of the allowed parameter space in the
large angle solution. To demonstrate the effect of the correlations,
we have calculated the allowed regions without the uncertainty
correlations, which is shown in Fig.~\ref{fig_MSW-uncertainties}(b).
The correlations are significant in the large-angle region where the
predicted Homestake rate is larger than its experimental central value, while
the Kamiokande rate is smaller than its experimental central value, and the
$^8$B flux uncertainty does not enlarge the allowed region
if the correlation between the two experiments are properly taken into
account.  Without the correlations, the allowed parameter space become
larger.  In Fig.~\ref{fig_MSW-uncertainties}(c) we display the allowed
region ignoring both the correlations and the Earth effect, and also using
the same experimental values used in Ref.~\cite{Krauss-Gates-White}.
The large-angle solution stretches to $\Delta m^2$ as small as
about $10^{-7} \, \mbox{eV}^2$ even at 90\% C.L.  Thus, the claim in
Ref~\cite{Krauss-Gates-White} that the allowed region is very large
was in fact due to their neglect of the correlations.

The allowed parameter space is significantly enlarged if the
Turck-Chi\`eze--Lopes SSM is used (Fig.~\ref{fig_TC-SSM}), which
predicts the smallest $^8$B flux and assigns the largest uncertainties
among the SSMs.  We have also carried out the MSW calculations with
doubled theoretical uncertainties [Fig.~\ref{fig_larger-errors}(a)],
and with a tripled the Homestake experimental uncertainty
[Fig.~\ref{fig_larger-errors}(b)].

The neutrino production distribution in the core and the electron
density distributions in the Sun and Earth are potential sources
of the uncertainties that are not included in the MSW calculations
above.  We have varied those quantities within reasonable range
and repeated the calculations, but no change is observed in the
combined allowed parameter space.

\subsection{%
	       Analytic Approximations in the MSW Calculations   %
}

Instead of solving the MSW differential equation numerically, we have
calculated the neutrino survival probability with an analytic approximation
proposed by Petcov \cite{Petcov} that is obtained from an exact solution of
the equation, assuming an exponential form of the electron density
distribution in the Sun, which is a good approximation except
for inside the core region ($ < 0.15 \times R_\odot$, where $R_\odot$ is
the solar radius).  The Parke formula
\cite{Parke}, another approximation that is the simplest and assumes
a linear electron density, yields essentially the same MSW plots as Petcov
formula except for $ \Delta m^2 / E \leq  3 \times 10^{-8} \mbox{eV}\, ^2 /
\mbox{MeV}$ [see Fig.~\ref{fig_MSW-approximations}(b)], where the density
variation is no longer well approximated with a linear function within
the neutrino oscillation length.  The adiabaticity
proposed by Pizzochero assuming the exponential electron density also gives a
good approximation except for $ \Delta m^2 / E \leq 7 \times 10^{-9}
\mbox{eV}\, ^2 / \mbox{MeV}$ [see Fig.~\ref{fig_MSW-approximations}(c)],
where the approximation fails because the oscillation length becomes
larger than the density scale height at large angles.
The three approximations are compared for the Kamiokande rate in
Fig.~\ref{fig_MSW-approximations}; differences are noticeable in the
large angle with small $\Delta m^2$.

One limitation of the Petcov formula (as well as the other two) is that
it fails to describe nonadiabatic level crossing when the neutrino
production point is at or close to the resonant point.
\footnote{%
An analytic prescription when the neutrino production site overlaps with
the resonance point is discussed by Haxton \cite{Haxton}.
}
As discussed by Krastev and Petcov \cite{Krastev-Petcov}, however, both
of the global solutions shown in
Fig.~\ref{fig_MSW-earth} are safe from this limitation.  For the $^7$Be and
$^8$B neutrinos with energies relevant for the experiments, the resonance
takes place at $\sim 0.15 R_\odot$ and $ \sim 0.4 R_\odot$, respectively,
while most of those neutrinos are produced within $ 0.1 R_\odot$.  The $pp$
neutrinos are potentially dangerous since their energy corresponds to the
resonance at or close to the center of the Sun, and there is a substantial
overlap with the production site; however, in the allowed
parameter space of the global fit, the MSW transitions of the $pp$ are
adiabatic and well described by the Petcov formula.  We estimate that the
uncertainty for the global fit due to the analytic approximation is $\leq$ 1\%.

\subsection{%
               Confidence Level Definitions  %
}

There are differences of the combined allowed MSW regions in the literature.
Especially the large-angle regions shown in
Refs.~\cite{Shi-Schramm,Krastev-Petcov,Krauss-Gates-White} are larger than our
calculation, and even a third solution is allowed at 90\% C.L. for
$\Delta m^2 \sim 10^{-7} \mbox{eV}\, ^2$ and $\sin^22\theta \sim 0.7$.  The
possible sources of the difference are: treatment of the SSM flux
uncertainties
and their correlations, the treatment of the Earth effect, the MSW analytic
approximations, and the experimental input data, some of which have
been discussed above.  The most significant difference, however, comes
from the statistical definition of the confidence levels, rather than
the details in MSW calculations.

In the joint $\chi^2$ analysis above, we have used
\begin{equation}
\label{eqn_chi-region}
	\chi^2 (\sin^22\theta, \Delta m^2)
                  \leq \chi^2_{\mbox{\scriptsize min}} + \Delta \chi^2
\end{equation}
with $\Delta \chi^2 = 6.0$ as the 95\% confidence level (C.L.) region
in the 2-dimensional $\log\sin^22\theta - \log\Delta m^2$ plane, where
$\chi^2_{\mbox{\scriptsize min}}$ is the global $\chi^2$ minimum.  This
definition assumes a gaussian distribution of the probability density around
the global minimum, which is only an approximation in our case, particularly
when we have multiple $\chi^2$ minima.  Another definition used in the
literature is to take the combined allowed regions simply as the overlap of
the different experiments as shown in Fig.~\ref{fig_MSW-overlap}. (See also
Ref.~\cite{Krastev-Petcov}.)  This definition can, however, overestimate the
allowed regions.  Consider, for example, the parameters that are marginally
allowed at 95\% C.L. by two experiments; if we take the overlap as the
combined fit, the parameters are allowed.  But the total $\chi^2$ can be
very large: $\chi^2$ should be about 12 ($= 2 \times 6.0$) at the edge of
the two allowed regions and is allowed only at 99.8\% C.L. by the $\chi^2$
analysis.  Taking the overlap of the allowed regions of the different
experiments displayed in Fig.~\ref{fig_MSW-overlap} clearly overestimates
the uncertainties when compared to the $\chi^2$ analysis shown in
Fig.~\ref{fig_MSW-earth}, especially for the large-angle solution.
Also the overlap procedure by definition ignores the correlations of
the uncertainties between different experiments.

Before improving the statistical definition of the uncertainties of the
obtained parameters, we consider the goodness-of-fit for each MSW solution
under the MSW hypothesis: if one of the MSW solutions is true, then how likely
is it to obtain the observed (or a larger) $\chi^2$?
The goodness-of-fit is calculated from the $\chi^2$ minimum for each MSW
solution.  Without the Earth effect, the $\chi^2$ minimum of the nonadiabatic
and large-angle solution are 0.5 and 4.9, respectively.  For 1 degree of
freedom (= 3 experiments -- 2 parameters), the probabilities of obtaining the
$\chi^2$ values equal to or larger than those $\chi^2$ values by chance
are 48\% and 3\%, respectively.  That is, the hypothesis that the
nonadiabatic solution is the true solution yields a good fit, while
the large-angle solution hypothesis is possible
statistically only at 3\%.  When the Earth effect and the Kamiokande II
day-night result (6 data points) are included, there is a third $\chi^2$
minimum in the large-angle region with $\sin^22\theta = 0.76$ and
$\Delta m^2 = 1.2 \times 10^{-7}\, \mbox{eV}\, ^2$, because of the
regeneration of the electron neutrinos in the Earth at night in the pp
and $^7$Be energy range.  The $\chi^2$ values are 3.1, 8.1, and 13.1
for the nonadiabatic, large-angle, and new large-angle solution,
respectively.  For 7 degrees of freedom (= 9 data -- 2 parameters),
the probabilities of getting $\chi^2$ larger than those values are 88,
32, and 7\%.  The fits for the first two solutions are reasonable, while
the third fit is somewhat poor but not excluded.  The small $\chi^2$
contribution from the Kamiokande day-night data is responsible for the
improvement of the large-angle, large $\Delta m^2$ solution.  The results are
summarized in Table~\ref{tab_goodness-noearth} and \ref{tab_goodness-earth}
and the corresponding results for sterile neutrinos in
Tables~\ref{tab_MSW-sterile-noearth} and \ref{tab_MSW-sterile}.

The nonadiabatic solution gives a better fit than the large-angle solutions,
either with or without the Earth effect, and this tendency can be quantified
in another way: suppose the two-flavor MSW is true and the probability density
of finding the true parameter is distributed throughout the $\log\sin^22\theta
- \log\Delta m^2$ plane, including all the $\chi^2$ minima. That is, we
assume that the probability distribution is
\begin{equation}
	P(\sin^22\theta, \Delta m^2) =
   			N \exp[ -\chi^2(\sin^22\theta, \Delta m^2)/2],
\end{equation}
where $N$ is chosen so that the total probability is unity.  Then what is the
relative probabilities of finding the true parameters in the different
regions?  We approximate the
probability distributions as an overlap of Gaussian distributions, each
corresponding to a different MSW region.  Then the relative probability of
finding the true parameters in the $i$th region ($i$ = nonadiabatic,
large-angle solutions) is calculated as
\begin{equation}
	P^i_{\mbox{\scriptsize relative}}
	= \frac{
                 2\pi \sigma^i_s \sigma^i_m
	              \sqrt{ 1 - \rho^{i 2} }
                      \exp( -\chi^2_{i,\mbox{\scriptsize min}}/2)
               }
               {
		      \sum_j
                 2\pi \sigma^j_s \sigma^j_m
	              \sqrt{ 1 - \rho^{j 2} }
                      \exp( -\chi^2_{j,\mbox{\scriptsize min}}/2)
               },
\end{equation}
where $\sigma_s$ and $\sigma_m$ are the standard deviations of
$\log\sin^22\theta$ and $\log\Delta m^2$, respectively, and $\rho^i$ is
the correlation
parameter; $j$ runs over each allowed region with the $\chi^2$ minimum
$\chi^2_{j,\mbox{\scriptsize min}}$.  Without the Earth effect, the relative
probabilities of finding the true parameters in the nonadiabatic and
large-angle region are 89\% and 11\%, respectively.  When the Earth effect
and the Kamiokande II day-night data are included, the probabilities for the
nonadiabatic and two large-angle solutions are 94.9\%, 4.6\%, and 0.5\%,
respectively.  The nonadiabatic solution is strongly favored.  The results are
summarized in Table~\ref{tab_goodness-noearth} and \ref{tab_goodness-earth}.

Finally, using the probability density approximated with multiple Gaussian
distributions, we improve the confidence level definition of the MSW parameter
uncertainties.  The confidence level $P$ and $\Delta \chi^2$ are related by
\begin{equation}
	P(\Delta \chi^2)
	= \frac{
		      \sum_i
                 2\pi \,
                      \sigma^i_s \sigma^i_m
	              \sqrt{ 1 - \rho^{i 2} } \,
                      {\cal P}_i(\Delta \chi^2)
               }
               {
		      \sum_j
                 2\pi \,
                      \sigma^j_s \sigma^j_m
	              \sqrt{ 1 - \rho^{j 2} }
                      \exp( -\chi^2_{j,\mbox{\scriptsize min}}/2)
               },
\end{equation}
where
\begin{equation}
	{\cal P}_i(\Delta \chi^2)
      = \int_{\chi^2_{i, \mbox{\scriptsize min}}}
            ^{\chi^2_{0, \mbox{\scriptsize min}} + \Delta \chi^2}
	\exp[ - \frac{1}{2}
                (  \chi^2_{0, \mbox{\scriptsize min}} + \chi^2
                 - \chi^2_{i, \mbox{\scriptsize min}}
                )
	    ] \;
              d\chi^2
\end{equation}
with
$	\chi^2_{0, \mbox{\scriptsize min}}
      = \min
                 \{ \,
	            \chi^2_{i, \mbox{\scriptsize min}} \, | \,
	            i = \mbox{all local minima}
                  \}.
$
Without the Earth effect, the 90, 95, and 99\% C.L. correspond to
$\Delta \chi^2 = 5.9, \; 7.3, \; \mbox{and} \; 10.6$, respectively.  With
the Earth effect, $\Delta \chi^2 = 5.5, \; 7.0, \; \mbox{and} \; 10.9$.
(These should be correspond with the values $\Delta \chi^2 = 4.6, \; 6.0,
\mbox{ and }9.2$
for a single Gaussian distribution.)  We believe that the new prescription
is more reliable because it takes into account the existence of multiple
minima and our lack of a priori knowledge of which is the true solution.
It is also a more conservative estimate of the uncertainties.
The allowed regions with the improved C.L. definition are shown in
Fig.~\ref{fig_MSW-newCL}.

\subsection{%
               Results of the Global MSW Analysis %
}

The results of the global MSW analysis for two-flavor oscillations
are displayed in Fig.~\ref{fig_MSW-best} and the best fit parameters are
listed in Tables~\ref{tab_goodness-earth}.  These include the SSM
uncertainties and their correlations, the Earth effect and the Kamiokande II
day-night data, the improved statistical definition of confidence
levels, and the updated experimental input data.
\footnote{%
There are additional Kamiokande II data of the energy spectrum of recoil
electrons \cite{Kamiokande}.  The spectrum slightly disfavors a part of
the adiabatic
region ($\Delta m^2 \sim 10^{-4}\; \mbox{eV}\,^2 \mbox{ and }
\sin^22\theta \sim 6 \times 10^{-3}$), but the effect in the combined fit
is insignificant.  The spectrum information is not statistically independent
and is therefore not included in our analysis.
}
Between the two allowed
regions at 90\% C.L., the nonadiabatic solution with
\begin{equation}
		\Delta m^2 \sim 6 \times 10^{-6} \, \mbox{eV}\, ^2
		\; \mbox{  and  } \;
		\sin^22\theta \sim 7 \times 10^{-3}
\end{equation}
gives an excellent description of the observations, while the
second (large-angle) solution with
\begin{equation}
		\Delta m^2 \sim 9 \times 10^{-6} \,\mbox{eV}\, ^2
		\; \mbox{  and  } \;
		\sin^22\theta \sim 0.6
\end{equation}
is marginally allowed at 90\% C.L.   In the second region the Earth
effect and the uncertainty correlations between the experiments are
significant: the regeneration of $\nu_e$
in the Earth during the night distorts and enlarges the allowed region
while the absence of the day-night asymmetry in the Kamiokande II data
excludes a wide parameter space \cite{HL,Earth-effect,Kam-daynight}.  The
omission of the correlation would
result in overestimating the uncertainties.  There is also a third solution
at large-angle and low $\Delta m^2$ that is allowed at 99\% C.L.\ when the
Earth effect, which in this region is significant for  the $pp$ and $^7$Be
fluxes, is included.

One can consider the MSW oscillation to sterile neutrinos instead of
$\nu_\mu$ or $\nu_\tau$.  In that case the
survival probability of electron neutrinos depends both on the
electron and neutron density in the Sun \cite{MSW,Bahcall-Ulrich,sterile}.
A more important difference is the lack of a neutral current contribution
in the Kamiokande experiment, which would amount to about 15\% of
the total signal for flavor oscillations, requiring a larger $\nu_e$
survival probability in the sterile neutrino case.
As a result the MSW effect now cannot completely resolve the larger
Kamiokande rate relative to that of Homestake, and the fit becomes poorer,
especially for the large-angle (with large $\Delta m^2$) solution.  There,
$\chi^2_{\mbox{\scriptsize min}}$ is 21.0 for 7 degrees of freedom
including the Earth effect, and no allowed region exists even at
99\% C.L. (see Fig.~\ref{fig_MSW-best-sterile}, \ref{fig_sterile},
Tables~\ref{tab_MSW-sterile-noearth}, and \ref{tab_MSW-sterile}.)
The large-angle region for sterile neutrinos is independently
excluded by the bound on the number of neutrino species in
big-bang nucleosynthesis \cite{sterile,Shi-Schramm-Fields}.

\section{%
              The MSW Effect in Nonstandard Solar Models             %
}
\label{sec_nonssms}

We have so far considered the MSW effect within the SSM uncertainties.  The
SSM is, however, still a theory that needs to be calibrated.  Although the
nonstandard solar models alone cannot explain the solar neutrino data
\cite{BHL}, it is possible that the MSW effect takes place while the SSM is
incorrect.  We have examined constraints on the MSW parameter space with
four different nonstandard solar models that are explicitly calculated
and have neutrino flux predictions. Each of these models is {\it ad hoc}
and assumes nonstandard input parameters that are grossly different from
those in the SSMs, addressing possibilities
of our ignorance of the astrophysical quantities such as the opacity or
the primordial element abundances, or of the nuclear reaction cross
sections that have never been measured at energies equivalent to
the solar temperature.  Two of them were constructed in
attempts to explain the solar neutrino deficit, and predict smaller
$^8$B and $^7$Be fluxes than the SSMs: the low opacity model by Dearborn
\cite{Dearborn}, in which the opacity is reduced by 20\% in the
region where the solar temperature is above $5 \times 10^6$ K, and the high
$S_{11}$ model with the $p+p$ cross section increased by 30\%
\cite{Castellani-etal}.  The allowed regions are generally enlarged
compared to the SSM case, and even the
adiabatic branch of $^8$B ($\Delta m^2 \sim 10^{-4} \;\mbox{eV}\,^2
\mbox{ and } \sin^22\theta \sim 10^{-4} - 1$) is allowed.  The MSW
parameters for the two models are shown in
Fig.~\ref{fig_MSW-nonssms}(a) and (b).
\footnote{%
No flux uncertainties are given in those models.  We used the standard
values of the uncertainties in Ref.~\cite{Bahcall-Pinsonneault}.}

The two other nonstandard solar models predict larger neutrino fluxes:
the high Y model \cite{Bahcall-Ulrich} that assumes a larger primordial
helium abundance in the solar interior, considered to solve problems with
helioseismology data, and the maximum rate model \cite{Bahcall-Pinsonneault},
in which $S_{33}$ is artificially set to zero, yielding the largest
prediction for the gallium rate (303 SNU).  With the larger fluxes the
combined allowed regions generally move inward in the MSW triangle, as
displayed in Fig.~\ref{fig_MSW-nonssms2}(a) and (b).

The solar models with a nonstandard opacity value or nonstandard heavy element
abundance are generally parametrized by a nonstandard central temperatures
($T_C$) \cite{BHL,BKL,BHKL}, and the neutrino fluxes are described by the
power laws of Eqn.~(\ref{eq_power-law}).  One can constrain the MSW parameter
space from the combined data for these generic nonstandard solar models
by allowing $T_C$ to change freely.  The allowed MSW regions are displayed in
Fig.~\ref{fig_MSW-Tcfree}.  The existing observations constrain
\begin{equation}
	T_C  =  1.02 \pm 0.02 \; (1 \sigma),
\end{equation}
with $\chi^2_{\mbox{\scriptsize min}} = 2.1$ for 6 degrees of
freedom (d.f.); the $T_C$ obtained by  observations allowing the MSW
effect is consistent with the SSM ($T_C = 1 \pm 0.006$).
We note that without the MSW effect there is no consistent $T_C$ to
describe the observations \cite{BHKL,BKL,BHL}.

Another major source of uncertainty in the SSM is the
$p + ^7\!\mbox{Be}$ cross section ($S_{17}$), which is poorly measured and has
the largest uncertainty among the SSM input parameters.  $S_{17}$ is
directly proportional to the $^8$B flux and is independent of the
astrophysical
uncertainties parametrized by $T_C$.  We have carried out MSW fits
assuming two extreme values of $S_{17}$; a 30\% reduction and a
50\% increase.  The results are shown in Fig.~\ref{fig_MSW-B8free}(a) and (b).
It is also reasonable to consider the $^8$B flux as a free parameter,
given the large uncertainties from both astrophysics and nuclear physics.
When the $^8$B flux is used as a completely free parameter in the MSW fit,
the experiments constrain
\begin{equation}
	\phi(\mbox{B}) / \phi(\mbox{B})_{\mbox{\scriptsize SSM}}
	= 1.43 + 0.65 - 0.42 \; (1 \sigma)
\end{equation}
with $\chi^2_{\mbox{\scriptsize min}} = 2.1$ for 6 d.f., and the allowed
regions are displayed in Fig.~\ref{fig_MSW-B8free}(c).
The larger values of $\phi(\mbox{B})$ and $T_C$ are preferred since a
larger $^8$B flux can reduce the relative difference in the survival
probabilities for Homestake and Kamiokande, giving more freedom in the
MSW parameter constraint.

\section{%
              MSW Predictions for Future Experiments %
}
\label{sec_future}

One can check the consistency between the current observations and the
MSW theory by predicting the gallium rate from the combined chlorine
and Kamiokande data \cite{Ga-prediction}.  The parameter space allowed by
the two experiments  is shown in Fig.~\ref{fig_Ga-prediction}(a), along with
the survival probability contours for the gallium experiment.
Assuming the Bahcall-Pinsonneault SSM, the combined
Homestake and Kamiokande observations predict the gallium rate to be
$ \leq 100$ SNU; the MSW prediction is consistent with the current
result of SAGE and GALLEX albeit the large theoretical uncertainty.
By reducing the statistical uncertainties, we estimate the allowed parameter
space expected in the near future.  Shown in Fig.~\ref{fig_Ga-prediction}(b) is
the region allowed by the combined observations when the gallium statistical
uncertainty is reduced by a factor $1/\sqrt{2}$, which is equivalent to the
data set of the gallium experiment by the end of 1994; the central value
is assumed to be the same as current data.  Although the nonadiabatic
region shrinks somewhat and the large-angle solution is no longer allowed at
95\% C.L., no significant change is expected unless the experimental
values change drastically.

Although the current observations are consistent with the two-flavor MSW
hypothesis and strongly disfavor the astrophysical solutions, a smoking
gun evidence for a nonstandard neutrino physics is still awaited.
Theoretical questions yet to be answered are: (a) distinguishing astrophysical
solutions and particle physics solutions (e.g., the MSW effect),
(b) calibrating solar models, and (c) distinguishing the MSW nonadiabatic
and large-angle solution and extracting $\Delta m^2$ and $\sin^22\theta$
from the data.

For the next-generation solar neutrino experiments, such as Sudbury Neutrino
Observatory (SNO) \cite{SNO}, Super-Kamiokande \cite{Super-Kamiokande},
BOREXINO \cite{BOREXINO}, and ICARUS \cite{ICARUS}, the MSW mechanism
can yield robust predictions and is a verifiable hypothesis.
The measurement of the charged to neutral current ratio (CC/NC) in SNO
and the measurements of the
neutrino energy spectrum and the day-night effect in SNO and Super-Kamiokande
should be able to answer the questions addressed above.

The measurement of the neutral current in SNO is insensitive to flavor
oscillations and is a direct measurement of the $^8$B flux.
One can calibrate the core temperature of the Sun at the
1\% level given the power law dependence of the flux ($\phi(\mbox{B})
\sim T_C^{18}$) and allowing a 20\% uncertainty from the
experimental error and the $S_{17}$ uncertainty.

The measurement of CC/NC in SNO would be the most direct test of the
MSW hypothesis.  The up-to-date global analysis predicts the ratio to be
\begin{equation}
		\mbox{CC/NC} =
			\left\{
			\begin{array}{ l l }
                        0.2 - 0.6  & \mbox{ (nonadiabatic solution) } \\
			0.2 - 0.3  & \mbox{ (large-angle solution) }
			\end{array}	\right.
\end{equation}
compared to the ratio expected if no oscillations occur.  This measurement
is, however, insensitive to oscillations between electron neutrinos and
sterile neutrinos.

The measurements of the charged current spectrum in SNO and the recoil
electron spectrum in Super-Kamiokande are important for distinguishing the
two solutions suggested by the global analysis of current observations.
Should the nonadiabatic solution, which gives the better fit for the data,
be the case, we expect a depletion of electron neutrinos in the lower end of
the observed spectrum, while little distortion is expected in the large-angle
solution (or for astrophysical solutions).   In Fig.~\ref{fig_SNO-CCspectrum},
we compare the spectra of the two solutions, one in the best fit solution
in the nonadiabatic region, and one in the large-angle region.  The
large-angle spectrum is normalized to the nonadiabatic spectrum above the
threshold.
The error bars indicate the statistical uncertainties assuming 6,000 events,
equivalent to two years of operation.  We have included in the calculation
the charged current cross section \cite{Nozawa,nu-d-cross-section}
and the detector energy resolution \cite{SNO}.  In
Fig.~\ref{fig_SuperK-spectrum}, the spectra of the two solutions are shown
for 16,000 events (two year operation) in Super-Kamiokande.  The spectrum
shape expected for oscillations to sterile neutrinos are almost identical
to these.

Another MSW prediction that helps to distinguish the two solutions is the
day-night effect due to matter oscillations in the Earth during the
night.  The effect shows up not only in day-night differences of signals,
but also in the time dependence during the night and in seasonal variations
due to the obliquity of the Earth.   The large-angle solutions are quite
sensitive to the Earth effect, but the nonadiabatic solution is insensitive
at the observable level.
\footnote{%
A possibility of detecting the day-night effect for the nonadiabatic
solution was discussed in Ref.~\cite{LoSecco}.  While it is
true that the instantaneous enhancement of the night-time signal can be
$\sim 20\%$ with respect to the day-time signal, it is hardly measurable
when the signals are averaged over time bins and the statistical uncertainties
are taken into account.  See Fig.~\ref{fig_SNO-day-night}(a) and
\ref{fig_SuperK-day-night}(a).
}
  The expected rates at night along with
the day-time rate are shown in Fig.~\ref{fig_SNO-day-night} and
\ref{fig_SuperK-day-night} for SNO and Super-Kamiokande, respectively.
The night rate is divided into six bins according to the angle between the
Sun and the nadir at the detectors, corresponding to bin neutrinos with
different path lengths in the Earth.  The effect yields such a noticeable
variation in the large-angle region that an updated Kamiokande day-night
data might be enough to confirm or rule out the large-angle solution.
The prediction for Kamiokande with error bars equivalent to 200 events
is shown in Fig.~\ref{fig_KIII-day-night}.

The spectrum distortions and the day-night effect are particularly
important for oscillations to sterile neutrinos, for which
the charged to neutral current ratio is unchanged by the oscillations and is
therefore the same as for astrophysical solutions.
The distortion of the spectrum signifies the nonadiabatic solution, while
the day-night effect indicates the large-angle solution, which is
already excluded at 99\% C.L. by the existing data.

When the SSM and two-flavor oscillations are assumed, one can predict
the rate in those experiments.  In Fig.~\ref{fig_future-detectors},
the combined allowed parameter space is displayed with the survival
probability contours of SNO (the charged current mode), Super-Kamiokande,
and BOREXINO.  Those high-counting experiments should be able to constrain
the parameter space precisely and check the consistency of the MSW
predictions.

\section{%
               Conclusions %
}

Various theoretical uncertainties in the global MSW analysis have been
discussed in detail.  It was shown that our parametrized SSM uncertainties
yield essentially the same result as the Monte Carlo estimation by
Bahcall and Ulrich.  The direct comparison was made in the flux
uncertainties and their correlations, in the rate uncertainties and
their correlations, and in the MSW calculations.    The different MSW
approximations of Petcov, Parke, and Pizzochero were compared.
Various confidence level definitions are discussed and a
careful statistical treatment in the global fit was emphasized.

There are two MSW global solutions at 90\% C.L., one in the nonadiabatic
region and the other in the large-angle, the former solution giving
a considerably better fit.  The proper treatment of the Earth effect
and the Kamiokande day-night data is significant in the large-angle region.
When the
Earth effect for the $^7$Be and $pp$ neutrinos are included, there is a
third solution at 99\% C.L in the large-angle, small $\Delta m^2$ region.
For the oscillations to sterile neutrinos, the solution is limited to
the nonadiabatic at 90\% C.L.  In Fig.~\ref{fig_big-picture} the MSW solutions
are displayed with other observational hints of neutrino mass: the
oscillation interpretation
of the atmospheric neutrinos \cite{atmospheric-neutrino-problem} and the
cold plus hot dark matter scenario to interpret the Cosmic Background
Explorer (COBE) microwave background anisotropy measurement and the
large-scale structure observations
\cite{CPHDM}. The see-saw predictions for neutrino mass and mixing are
also displayed in Fig.~\ref{fig_big-picture}: $\nu_e \leftrightarrow \nu_\mu $
oscillations in the $SO_{10}$ grand unified theory (GUT) with an
intermediate-scale breaking \cite{BKL,Langacker-Neutrino-Telescopes},
$\nu_e \leftrightarrow \nu_\tau$ oscillations in the supersymmetric $SO_{10}$
GUT \cite{BKL,Langacker-Neutrino-Telescopes}, and
$\nu_e \leftrightarrow \nu_\mu$ oscillations in the superstring-inspired
model with nonrenormalizable operators \cite{Cvetic-Langacker}.  In the
$SO_{10}$ GUT, the $\nu_\tau$ mass is expected to be in a range relevant
to the cosmological hot dark matter.  Assuming the Bahcall-Pinsonneault
SSM, the MSW
solutions for $\Delta m^2$ are generally in agreement with the theoretical
expectations, especially with the $\nu_e \leftrightarrow \nu_\tau$
oscillations in the supersymmetric $SO_{10}$ model.  However, the mixing
angles are not consistent with the expectation of the simplest versions
of the models that the lepton mixing angles are similar to the corresponding
quark mixing angles \cite{BKL}.  (For the string-inspired model of
Ref.~\cite{Cvetic-Langacker} there is no compelling prediction for the
mixing angles.)

The global MSW was also carried out for various standard and
nonstandard solar models; with the Turck-Chi\`eze--Lopes SSM or nonstandard
solar models a wider range of the parameter space is possible.  When
the core temperature and the $^8$B flux each was used as a free fitting
parameter, the data constrained $T_C = 1.02 \pm 0.02$ and
$\phi(\mbox{B})/\phi(\mbox{B})_{\mbox{\scriptsize SSM}} =
1.43 + 0.65 - 0.42$ at $1 \sigma$, respectively.

The predictions of the MSW solutions assuming the Bahcall-Pinsonneault SSM
were discussed in detail for SNO and Super-Kamiokande.  We expect for
flavor oscillations the charged to neutral current ratio
in SNO to be
0.2 -- 0.6 and 0.2 -- 0.3 of the SSM prediction for the nonadiabatic
and large-angle solution, respectively.  The nonadiabatic solution yields
spectrum distortions measurable in SNO and Super-Kamiokande, while the
large-angle solution predicts characteristic day-night differences.
The spectrum distortion and the day-night effect are independent of
solar models, and  are particularly important for the oscillations to
sterile neutrinos because the absence of the neutral current in SNO
prevents one from distinguishing neutrino oscillations from astrophysical
solutions in the charged to neutral current ratio measurement.

\acknowledgments                                              %**revtex**

It is pleasure to thank Eugene Beier and Sidney Bludman for useful
discussions.  John Bahcall kindly provided us the Monte Carlo SSM data
file.  We thank David Dearborn for the data of the low opacity models,
and Serguey Petcov for a useful correspondence concerning the analytic
MSW approximations.  We are grateful to Satoshi Nozawa for providing us
a neutrino-deuteron cross section code.  This work is supported by the
Department of Energy Contract No.\ DE-AC02-76-ERO-3071.

% The end of text ************************************************************

% Here comes the reference ****************************************************

%                        * * * REFERENCE * * *

% Here comes the tables *******************************************************

%\onecolumn                                                       %**2column**
\begin{table}[p]
\caption{
%
%                                  TABLE I
%
The standard solar model predictions of Bahcall and Pinsonneault
\protect\cite{Bahcall-Pinsonneault} (BP SSM) and of Turck-Chi\`eze
and Lopes \protect\cite{Turck-Chieze-Lopes} (TCL SSM), along with
the results of the solar neutrino experiments.  The gallium experiment
is the combined result of SAGE and GALLEX I and II.
}
\label{tab_exps}
\vspace{1.0ex}
\begin{tabular}{ l  c c c }
%
%------------------------------------------------------------------------------
%------------------------------------------------------------------------------
               & BP SSM        & TCL SSM      & Experiments \\
\hline%------------------------------------------------------------------------
Kamiokande \tablenotemark[1]
               &  1 $\pm$ 0.14 & 0.77 $\pm$ 0.19 & 0.51 $\pm$ 0.07 BP SSM  \\
Homestake \tablenotemark[2] (SNU)
               &  $8\pm 1$   & 6.4 $\pm$ 1.4  & 2.32 $\pm$ 0.23
                                                (0.29 $\pm$ 0.03 BP SSM) \\
SAGE \tablenotemark[3] \& GALLEX \tablenotemark[4] (SNU)
               & 131.5 $^{+7}_{-6}$ & 122.5 $\pm$ 7 & 81 $\pm$ 13
                                                    (0.62 $\pm$ 0.10 BP SSM)
%------------------------------------------------------------------------------
%------------------------------------------------------------------------------
\tablenotetext[1]{%
The result of the combined data of 1040 days of Kamiokande II
[0.47 $\pm$ 0.05 (stat) $\pm$ 0.06 (sys) BP SSM]
and 514.5 days of Kamiokande III
[0.57 $\pm$ 0.06 (stat) $\pm$ 0.06 (sys) BP SSM]
\cite{Kamiokande-update}.
}
\tablenotetext[2]{%
The result of Run 10 to 124 (through May, 1993) \cite{Homestake-update}.
}
\tablenotetext[3]{%
The preliminary result of SAGE I (from January, 1990 through May, 1992) is
70 $\pm$ 19 (stat) $\pm$ 10 (sys) SNU \cite{SAGE-update}.
}
\tablenotetext[4]{%
The combined result of GALLEX I and II (including 21 runs through April, 1993)
is 87 $\pm$ 14 (stat) $\pm$ 7 (sys) SNU \cite{GALLEX}.
}

\end{tabular}
\end{table}

%\twocolumn                                                       %**2column**
\begin{table}[p]
\caption{
%
%                                  TABLE II
%
The parameters of the SSM flux uncertainties.  $n$ is the
exponent in the $T_C$ power law, and $s_k = \Delta S_k / S_k$ ($k = 34, 17$)
are the fractional flux uncertainties due to the nuclear reaction cross
section $^3\mbox{He}+^4\!\mbox{He}$ and $p + ^7\!\mbox{Be}$.  We quote
$s_k$ from Ref.~\protect\cite{Bahcall-Ulrich,Bahcall-Pinsonneault}.
We determine $\Delta T_C$ to be 0.0057 for Bahcall-Ulrich SSM and
0.0060 for the Bahcall-Pinsonneault SSM.
}
\label{tab_parameters}
\vspace{1.0ex}
\begin{tabular}{ l  c c c }
%
%------------------------------------------------------------------------------
%------------------------------------------------------------------------------
               & n      & $s_{34}$ & $s_{17}$ \\
\hline%------------------------------------------------------------------------
\multicolumn{1}{l}{Bahcall-Ulrich SSM ($\Delta T_C = 0.0057$)}  \\
$pp$           & $-$1.2 & 0       & 0              \\
$^7$Be         & 8      & 0.02    & 0              \\
$^8$B          & 18     & 0.02    & 0.07           \\
\hline%------------------------------------------------------------------------
\multicolumn{1}{l}{Bahcall-Pinsonneault SSM ($\Delta T_C = 0.0060$)}  \\
$pp$           & $-$1.2 & 0       & 0                \\
$^7$Be         & 8      & 0.032   & 0                \\
$^8$B          & 18     & 0.032   & 0.093            \\
%------------------------------------------------------------------------------
%------------------------------------------------------------------------------
%
\end{tabular}
\end{table}

\clearpage                                                        %**2column**
\begin{table}[p]
\caption{
%
%                                  TABLE III
%
The magnitudes of flux uncertainties ($\Delta \phi / \phi$ at $1 \sigma$)
quoted from the Bahcall-Ulrich SSM \protect\cite{Bahcall-Ulrich},
of the Bahcall-Ulrich Monte Carlo SSMs (Gaussian fit), and of the
parametrized Bahcall-Ulrich SSM using the central temperature and
the nuclear reaction cross sections.  Also listed are the
uncertainties of the Bahcall-Pinsonneault SSM and of its parametrized fluxes.
(The Monte Carlo study of the Bahcall-Pinsonneault model is not available.)
}
\label{tab_magnitudes}
\vspace{1.0ex}
\begin{tabular}{ l  c c c c }
%
%------------------------------------------------------------------------------
%------------------------------------------------------------------------------
                                     &    $pp$     &   $^7$Be     &  $^8$B   \\
\hline%------------------------------------------------------------------------
Bahcall-Ulrich SSM                   &  0.0059     &   0.050      &  0.12   \\
Bahcall-Ulrich SSM (Monte Carlo)     &  0.0067     &   0.05       &  0.17   \\
Parametrized ($\Delta T_C = 0.0057$) &  0.0069     &   0.05       &  0.13   \\
\hline%------------------------------------------------------------------------
Bahcall-Pinsonneault SSM             &  0.0067     &   0.06       &  0.14   \\
Parametrized ($\Delta T_C = 0.0060$) &  0.007      &   0.06       &  0.15   \\
%------------------------------------------------------------------------------
%------------------------------------------------------------------------------
%
\end{tabular}
\end{table}

\vspace{10ex}
\begin{table}[p]
\caption{
%
%                                  TABLE IV
%
The correlation matrices of flux uncertainties obtained from the Bahcall-Ulrich
Monte Carlo SSMs and the parameterization method.  The agreement between
the two methods is good, especially for the $pp$ -- $^7$Be element.
}
\label{tab_correlations}
\vspace{1.0ex}
\begin{tabular}{ l  c c c c }
%
%------------------------------------------------------------------------------
%------------------------------------------------------------------------------
                                     &    $pp$     &  $^8$B       &  $^7$Be  \\
\hline%------------------------------------------------------------------------
\multicolumn{1}{l}{Bahcall-Ulrich SSM (Monte Carlo)}     &&                \\
$pp$                                 &   1         &              &         \\
$^8$B                                &$-$0.73      &   1          &         \\
$^7$Be                               &$-$0.92      &   0.74       &   1     \\
\hline%------------------------------------------------------------------------
\multicolumn{3}{l}{Parametrized with $\Delta T_C$ and $\Delta s$}           \\
$pp$                                 &   1         &              &         \\
$^8$B                                &$-$0.81      &   1          &         \\
$^7$Be                               &$-$0.92      &   0.80       &   1     \\
%------------------------------------------------------------------------------
%------------------------------------------------------------------------------
%
\end{tabular}
\end{table}

\begin{table}[p]
\caption{
%
%                                  TABLE V
%
The fractional contribution from flux components for the predicted SSM rates
for the different solar neutrino detectors.  The total SSM value is normalized
to one.
}
\label{tab_fractions}
\vspace{1.0ex}
\begin{tabular}{ l  c c c }
%
%------------------------------------------------------------------------------
%------------------------------------------------------------------------------
	            & Kamiokande      &  Cl       &  Ga                  \\
\hline%------------------------------------------------------------------------
\multicolumn{4}{l}{Bahcall-Ulrich SSM}                                    \\
$pp$                & 0               &  0        &    0.536          \\
$^7$Be (I)          & 0               &  0        &    0.009          \\
$^7$Be (II)         & 0               &  0.139    &    0.251          \\
$^8$B               & 1               &  0.772    &    0.106          \\
$pep$               & 0               &  0.025    &    0.023          \\
$^{13}$N            & 0               &  0.013    &    0.029          \\
$^{15}$O            & 0               &  0.038    &    0.046          \\
\hline%------------------------------------------------------------------------
\multicolumn{4}{l}{Bahcall-Pinsonneault SSM}                              \\
$pp$                & 0               &  0        &    0.538          \\
$^7$Be (I)          & 0               &  0        &    0.009          \\
$^7$Be (II)         & 0               &  0.150    &    0.264          \\
$^8$B               & 1               &  0.775    &    0.105          \\
$pep$               & 0               &  0.025    &    0.024          \\
$^{13}$N            & 0               &  0.013    &    0.023          \\
$^{15}$O            & 0               &  0.038    &    0.037          \\
%------------------------------------------------------------------------------
%------------------------------------------------------------------------------
%
\end{tabular}
\end{table}

\begin{table}[p]
\caption{
%
%                                  TABLE VI
%
The comparison of the magnitudes of rate uncertainties for the Bahcall-Ulrich
SSM, Monte Carlo SSMs, and the parametrized SSM.
Also listed are the uncertainties in the Bahcall-Pinsonneault SSM and its
parametrized SSM.  The detector cross section uncertainties are included
in the chlorine (Cl) and gallium (Ga) uncertainties.
}
\label{tab_exp_magnitudes}
\vspace{1.0ex}
\begin{tabular}{ l  c c c c }
%
%------------------------------------------------------------------------------
%------------------------------------------------------------------------------
                                     &  Kamiokande &      Cl      &  Ga   \\
\hline%------------------------------------------------------------------------
Bahcall-Ulrich SSM                   &  0.12       &  0.11     &$+0.05 -0.04$\\
Monte Carlo                          &  0.12       &  0.11        &   0.05  \\
Parametrized ($\Delta T_C = 0.0057$) &  0.13       &  0.11        &   0.05  \\
\hline%------------------------------------------------------------------------
Bahcall-Pinsonneault SSM             &  0.14       &   0.13       &  0.05   \\
Parametrized ($\Delta T_C = 0.0060$) &  0.15       &   0.13       &  0.05   \\
%------------------------------------------------------------------------------
%------------------------------------------------------------------------------
%
\end{tabular}
\end{table}

%\newpage
\vspace{10ex}
\begin{table}[p]
\caption{
%
%                                  TABLE VII
%
The comparison of the correlation matrices of rate uncertainties for
the Bahcall-Ulrich Monte Carlo SSMs and the parametrized SSM.
The agreement is excellent.
}
\label{tab_exp_correlations}
\vspace{1.0ex}
\begin{tabular}{ l  c c c c }
%
%------------------------------------------------------------------------------
%------------------------------------------------------------------------------
                                  &    Kamiokande     &   Cl     &  Ga   \\
\hline%------------------------------------------------------------------------
\multicolumn{1}{l}{Bahcall-Ulrich SSM (Monte Carlo)}                       \\
Kamiokande                        &   1         &              &         \\
Cl                                &$-$0.997     &   1          &         \\
Ga                                &$-$0.92      &   0.95       &   1     \\
\hline%------------------------------------------------------------------------
\multicolumn{1}{l}{Parametrized with $\Delta T_C$ and $\Delta s$}           \\
Kamiokande                        &   1         &              &         \\
Cl                                &$-$0.996     &   1          &         \\
Ga                                &$-$0.90      &   0.94       &   1     \\
%------------------------------------------------------------------------------
%------------------------------------------------------------------------------
%
\end{tabular}
\end{table}

\begin{table}[p]
\caption{
%
%                                  TABLE VIII
%
The best fit values of the MSW parameters when the Earth effect is ignored.
$P$ is the goodness-of-fit, i.e.,  the probability of obtaining by chance
a $\chi^2$ equal to or larger
than the obtained $\chi^2$. $P_{\mbox{\protect\scriptsize relative}}$ is
the relative probability between the different solutions when the
probability distribution is gaussian for each solution.  The improved
definition of 90, 95, and 99\% C.L. correspond to $\Delta \chi^2$ =
5.9, 7.3, and 10.6 [see Fig.~\protect\ref{fig_MSW-newCL}(a)].
}
\label{tab_goodness-noearth}
\vspace{1.0ex}
\begin{tabular}{ l  c c  }
%
%------------------------------------------------------------------------------
%------------------------------------------------------------------------------
		  &	Nonadiabatic  &  Large Angle I  \\
\hline%------------------------------------------------------------------------
  $\sin^22\theta$    & $6.4\times 10^{-3}$ & 0.71      \\
  $\Delta m^2$ (eV$^2$)& $6.3\times 10^{-6}$ & $1.6 \times 10^{-5}$ \\
\hline%------------------------------------------------------------------------
  $\chi^2$ (1 d.f.) &  0.5              & 4.9             \\
  $P$ (\%)          &  48               & 3        	\\
  $P_{\mbox{\scriptsize relative}}$ (\%)
                    &  89     	      & 11		\\
%
%------------------------------------------------------------------------------
%------------------------------------------------------------------------------
%
\end{tabular}
\end{table}

%\newpage
\vspace{10ex}
\begin{table}[p]
\caption{
%
%                                  TABLE IX
%
Same as Table~\protect\ref{tab_goodness-noearth} except that the Earth effect
and the Kamiokande II day-night effect are included.  The improved definition
of 90, 95, and 99\% C.L. correspond to $\Delta \chi^2$ = 5.5, 7.0, and 10.9
[see Fig.~\protect\ref{fig_MSW-newCL}(b)].
}
\label{tab_goodness-earth}
\vspace{1.0ex}
\begin{tabular}{ l  c c c }
%
%------------------------------------------------------------------------------
%------------------------------------------------------------------------------
		  &	Nonadiabatic  &  Large Angle I  & Large Angle II \\
\hline%------------------------------------------------------------------------
  $\sin^22\theta$    & $6.5\times 10^{-3}$ & 0.62      & 0.76          \\
  $\Delta m^2$ (eV$^2$)& $6.1\times 10^{-6}$ & $9.4\times 10^{-6}$
                                                   & $1.2\times 10^{-7}$\\
\hline%------------------------------------------------------------------------
  $\chi^2$ (7 d.f.) &  3.1              & 8.1		& 13.1             \\
  $P$ (\%)        &  88               & 32		& 7               \\
  $P_{\mbox{\scriptsize relative}}$ (\%)
                  &  94.9     	      & 4.6             & 0.5              \\
%
%------------------------------------------------------------------------------
%------------------------------------------------------------------------------
%
\end{tabular}
\end{table}

\newpage
\begin{table}[p]
\caption{
%
%                                  TABLE X
%
Same as Table~\protect\ref{tab_goodness-noearth} except that the oscillations
are for sterile neutrinos.  The Earth effect and the Kamiokande II
day-night effect are not included.  The improved definition of 90, 95, and
99\% C.L. correspond to $\Delta \chi^2$ = 4.8, 6.5, and 10.6.
}
\label{tab_MSW-sterile-noearth}
\vspace{1.0ex}
\begin{tabular}{ l  c c  }
%
%------------------------------------------------------------------------------
%------------------------------------------------------------------------------
		  &	Nonadiabatic  &  Large Angle I  \\
\hline%------------------------------------------------------------------------
  $\sin^22\theta$    & $7.4\times 10^{-3}$ & 0.85      \\
  $\Delta m^2$ (eV$^2$)& $4.7\times 10^{-6}$ & $9.6 \times 10^{-6}$ \\
\hline%------------------------------------------------------------------------
  $\chi^2$ (1 d.f.)&  2.8              & 11.7              \\
  $P$ (\%)        &  9.4              & 0.06        	\\
  $P_{\mbox{\scriptsize relative}}$ (\%)
                  &  98.9     	      & 1.1		\\
%
%------------------------------------------------------------------------------
%------------------------------------------------------------------------------
%
\end{tabular}
\end{table}

%\newpage
\vspace{10ex}
\begin{table}[p]
\caption{
%
%                                  TABLE XI
%
Same as Table~\protect\ref{tab_MSW-sterile-noearth} except that the
Earth effect and the Kamiokande II day-night effect are included.
The improved definition of 90, 95, and 99\% C.L. correspond to $\Delta
\chi^2$ = 4.6, 6.0, and 9.1.
}
\label{tab_MSW-sterile}
\vspace{1.0ex}
\begin{tabular}{ l  c c c }
%
%------------------------------------------------------------------------------
%------------------------------------------------------------------------------
		  &	Nonadiabatic  &  Large Angle I  & Large Angle II \\
\hline%------------------------------------------------------------------------
  $\sin^22\theta$    & $7.0\times 10^{-3}$ & 0.77      & 0.60          \\
  $\Delta m^2$ (eV$^2)$& $4.5\times 10^{-6}$ & $6.7\times 10^{-6}$
                                                   & $6.9\times 10^{-8}$\\
\hline%------------------------------------------------------------------------
  $\chi^2$ (7 d.f.)&  7.0              & 21.0		& 24.2             \\
  $P$ (\%)        &  43               & 0.4		& 0.1               \\
  $P_{\mbox{\scriptsize relative}}$ (\%)
                  &  99.59     	      & 0.03            & 0.01             \\
%
%------------------------------------------------------------------------------
%------------------------------------------------------------------------------
%
\end{tabular}
\end{table}

%\onecolumn                                                        %**2column**
% Here comes the figures ******************************************************
%
%                                  FIGURE 1
%
\begin{figure}[t]

%\vspace*{4ex}                                                        %**epsf**
%\begin{center}                                                       %**epsf**
%\begin{tabular}{c}                                                   %**epsf**
%\setlength{\epsfxsize}{\histsize}                                    %**epsf**
%\subfigure[$pp$]{\epsfbox{figures/mcflux_pp.ps}}                     %**epsf**
%\subfigure[$^7\mbox{Be}$]{\epsfbox{figures/mcflux_be7.ps}}           %**epsf**
%\subfigure[$^8$B]{\epsfbox{figures/mcflux_b8.ps}}                    %**epsf**
%\end{tabular}                                                        %**epsf**
%\end{center}                                                         %**epsf**
%
\caption{
The distributions of the (a) $pp$, (b) $^7$Be, and (c) $^8$B flux of the
1000 Monte Carlo SSMs by Bahcall-Ulrich (histograms) are fit with the
parametrized method (solid curves) that assumes Gaussian distributions
of the central temperatures and the nuclear reaction cross sections
around their central values.
}
\label{fig_gaussian_fluxes}
\end{figure}

%\clearpage                                                           %**epsf**
%
%                                  FIGURE 2
%
\begin{figure}[p]
%
%\vspace*{4ex}                                                        %**epsf**
%\begin{center}                                                       %**epsf**
%\begin{tabular}{c}                                                   %**epsf**
%\setlength{\epsfxsize}{\histsize}                                    %**epsf**
%\subfigure[Kamiokande ($^8$B)]{\epsfbox{figures/mcrate_kam.ps}}      %**epsf**
%\subfigure[Chlorine]{\epsfbox{figures/mcrate_cl.ps}}                 %**epsf**
%\subfigure[Gallium]{\epsfbox{figures/mcrate_ga.ps}}                  %**epsf**
%\end{tabular}                                                        %**epsf**
%\end{center}                                                         %**epsf**
%
\caption{
The distributions of the experimental rates obtained from Bahcall-Ulrich
Monte Carlo fluxes for the (a) Kamiokande, (b) chlorine, and (c)
gallium experiments (histograms).  They are compared with the rate
distributions obtained from the parametrized fluxes (solid curve).
In both cases the detector cross section uncertainties are not included.
}
\label{fig_gaussian_exps}
\end{figure}

%\clearpage                                                           %**epsf**
%
%                                  FIGURE 3
%
%\vspace*{-14ex}                                                      %**epsf**
\begin{figure}[p]
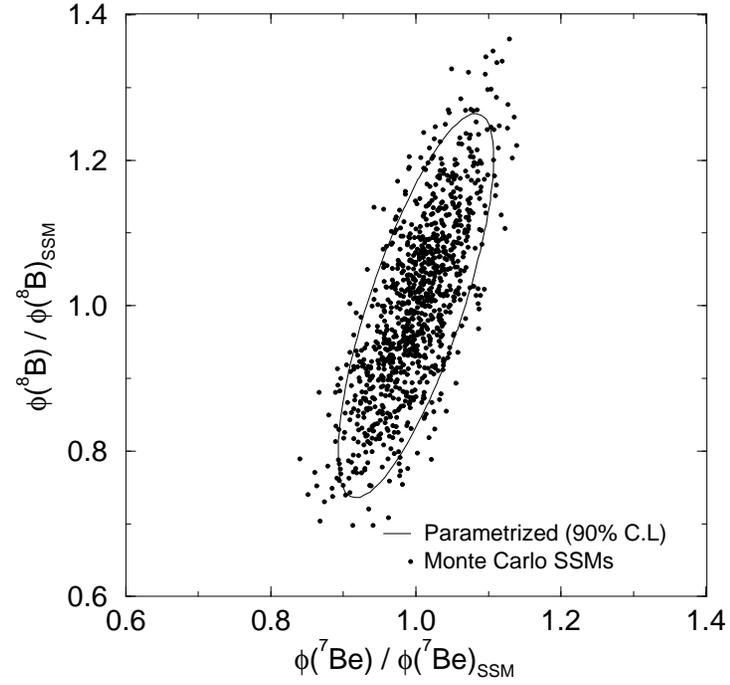

%
%\postscript{figures/bu_BeB.ps}{\scatsize}                            %**epsf**
%\vspace{-2cm}                                                        %**epsf**
%
\caption{
The distributions of the $^7$Be and $^8$B flux of the Bahcall-Ulrich SSMs
(dots), and the 90\% C.L. contour of our parametrized SSM (solid curve).
The magnitudes and the correlations of the fluxes are in excellent
agreement for the two methods.
}
\label{fig_correlations}
\end{figure}

%\clearpage                                                           %**epsf**
%
%                                  FIGURE 4
%
\begin{figure}[p]
%
%\begin{center}                                                       %**epsf**
%\begin{tabular}{c}                                                   %**epsf**
%\setlength{\epsfxsize}{\mswsize}                                     %**epsf**
%\subfigure[Monte Carlo SSM]{\epsfbox{figures/p_comb_bumc_noearth.ps}}%**epsf**
%\subfigure[Bahcall-Ulrich SSM]                                       %**epsf**
%                      {\epsfbox{figures/p_comb_bussm_noearth.ps}}    %**epsf**
%\subfigure[Bahcall-Pinsonneault SSM]                                 %**epsf**
%                       {\epsfbox{figures/p_comb_bpssm_noearth.ps}}   %**epsf**
%\end{tabular}                                                        %**epsf**
%\end{center}                                                         %**epsf**
%
\caption{
The allowed regions of the Homestake, Kamiokande, and
gallium experiments, and the combined observations.  The Earth effect
is not included.  The SSM uncertainties are calculated by (a) the
Bahcall-Ulrich 1000 Monte Carlo SSMs, (b) the parametrized
Bahcall-Ulrich SSM, and (c) the parametrized Bahcall-Pinsonneault SSM.
The comparison of (a) and (b) shows that the Monte Carlo method and our
parametrization method are essentially equivalent.  The allowed regions
are defined by $\chi^2(\sin^22\theta, \Delta m^2) \leq
\chi^2_{\mbox{\protect\scriptsize min}} + \Delta \chi^2$ with $\Delta \chi^2
= 6.0$ at 95\% C.L.
}
\label{fig_MSW-noearth}
\end{figure}

%\clearpage                                                           %**epsf**
%
%                                  FIGURE 5
%
\begin{figure}[p]
%
%\begin{center}                                                       %**epsf**
%\begin{tabular}{c}                                                   %**epsf**
%\setlength{\epsfxsize}{\mswsize}                                     %**epsf**
%\subfigure[Monte Carlo SSM]{\epsfbox{figures/p_comb_bumc.ps}}        %**epsf**
%\subfigure[Bahcall-Ulrich SSM]{\epsfbox{figures/p_comb_bussm.ps}}    %**epsf**
%\subfigure[Bahcall-Pinsonneault SSM]                                 %**epsf**
%                              {\epsfbox{figures/p_comb_bpssm.ps}}    %**epsf**
%\end{tabular}                                                        %**epsf**
%\end{center}                                                         %**epsf**
%
\caption{
Same as Fig.~\protect\ref{fig_MSW-noearth}, except that the Earth effect is
included.  The Kamiokande day-night data exclude the region shown with
a dotted line (90\% C.L.), and are incorporated in the combined fit.
}
\label{fig_MSW-earth}
\end{figure}

%\clearpage                                                           %**epsf**
%
%                                  FIGURE
%
\begin{figure}[p]
%
%\begin{center}                                                       %**epsf**
%\begin{tabular}{c}                                                   %**epsf**
%\setlength{\epsfxsize}{\mswsize}                                     %**epsf**
%\subfigure[No theory errors]{\epsfbox{figures/p_comb_noth-error.ps}} %**epsf**
%\subfigure[No correlations of theory errors]                         %**epsf**
%                            {\epsfbox{figures/p_comb_nocorrel.ps}}   %**epsf**
%\subfigure[No correlations without the Earth effect]                 %**epsf**
%      {\epsfbox{figures/p_comb_krauss_noearth.ps}}                   %**epsf**
%\end{tabular}                                                        %**epsf**
%\end{center}                                                         %**epsf**
%
\caption{
The experimental constraints on the MSW parameters assuming the
Bahcall-Pinsonneault SSM when (a) the theoretical
uncertainties of the SSM fluxes and the detector cross sections are ignored,
(b) the theoretical uncertainties are included, but their correlations
between the fluxes and between the experiments are ignored.  Displayed in
(c) is the result at 90\% C.L. calculated with the same condition in
Ref~\protect\cite{Krauss-Gates-White}: the correlations and the Earth
effect are ignored and the same experimental input parameters are used.
The omission of the correlations overestimates the allowed large-angle
regions.
}
\label{fig_MSW-uncertainties}
\end{figure}

%\clearpage                                                           %**epsf**
%
%                                  FIGURE
%
\begin{figure}[p]
%
%\begin{center}                                                      %**epsf**
%\begin{tabular}{c}                                                  %**epsf**
%\setlength{\epsfxsize}{\mswsize}                                    %**epsf**
%\subfigure[Without the Earth effect]                                %**epsf**
%                      {\epsfbox{figures/p_comb_tcssm_noearth.ps}}   %**epsf**
%\subfigure[With the Earth effect]{\epsfbox{figures/p_comb_tcssm.ps}}%**epsf**
%\end{tabular}                                                       %**epsf**
%\end{center}                                                        %**epsf**
%
\caption{
The experimental constraints when the Turck-Chi\`eze--Lopes SSM is
assumed.  The two figures are (a) without and (b) with the Earth
effect.  Since this SSM predicts a smaller $^8$B flux with a
larger uncertainty compared to the Bahcall-Pinsonneault SSM, the
allowed regions are noticeably larger than those shown in
Fig.~\protect\ref{fig_MSW-noearth} and \protect\ref{fig_MSW-earth}.
Especially, no constraints are obtained at 95\% C.L. from the upper
limit of the the Kamiokande result, and the Kamiokande allowed
regions are outside the triangles with the solid lines.  The region
excluded by the Kamiokande II day-night data is shown with the
dotted line (90\% C.L.) in (b).
}
\label{fig_TC-SSM}
\end{figure}

%\clearpage                                                           %**epsf**
%
%                                  FIGURE
%
\begin{figure}[p]
%
%\begin{center}                                                      %**epsf**
%\begin{tabular}{c}                                                  %**epsf**
%\setlength{\epsfxsize}{\mswsize}                                    %**epsf**
%\subfigure[Theory errors doubled]                                   %**epsf**
%                      {\epsfbox{figures/p_comb_th-errorx2.ps}}      %**epsf**
%\subfigure[Homestake error tripled]                                 %**epsf**
%              {\epsfbox{figures/p_comb_cl-errorx3.ps}}              %**epsf**
%\end{tabular}                                                       %**epsf**
%\end{center}                                                        %**epsf**
%
\caption{
The experimental constraints (a) when theoretical uncertainties of the
Bahcall-Pinsonneault SSM fluxes and the detector cross sections are
doubled, and (b) the
Homestake experimental uncertainty is tripled.  The Earth effect and
the Kamiokande II day-night data are included in both cases.
}
\label{fig_larger-errors}
\end{figure}

%\clearpage                                                           %**epsf**
%
%                                  FIGURE
%
\begin{figure}[p]
%
%\begin{center}                                                       %**epsf**
%\begin{tabular}{c}                                                   %**epsf**
%\setlength{\epsfxsize}{\mswsize}                                     %**epsf**
%\subfigure[Petcov formula]{\epsfbox{figures/c_kam_petcov.ps}}        %**epsf**
%\subfigure[Parke formula]{\epsfbox{figures/c_kam_lz.ps}}             %**epsf**
%\subfigure[Pizzochero formula]                                       %**epsf**
%                      {\epsfbox{figures/c_kam_pizzochero.ps}}        %**epsf**
%\end{tabular}                                                        %**epsf**
%\end{center}                                                         %**epsf**
%
\caption{
The Kamiokande contours of the signal to SSM ratio (including the neutral
current reaction in flavor oscillations) for three different
analytic MSW approximations of (a) Petcov, (b) Parke, and (c) Pizzochero.
The formulae by Parke and Pizzochero fail in the large-angle,
small $\Delta m^2$ region ($\Delta m^2 \leq 3 \times 10^{-8} \mbox{eV}\, ^2$).
}
\label{fig_MSW-approximations}
\end{figure}

%\clearpage                                                           %**epsf**
%
%                                  FIGURE
%
\begin{figure}[p]
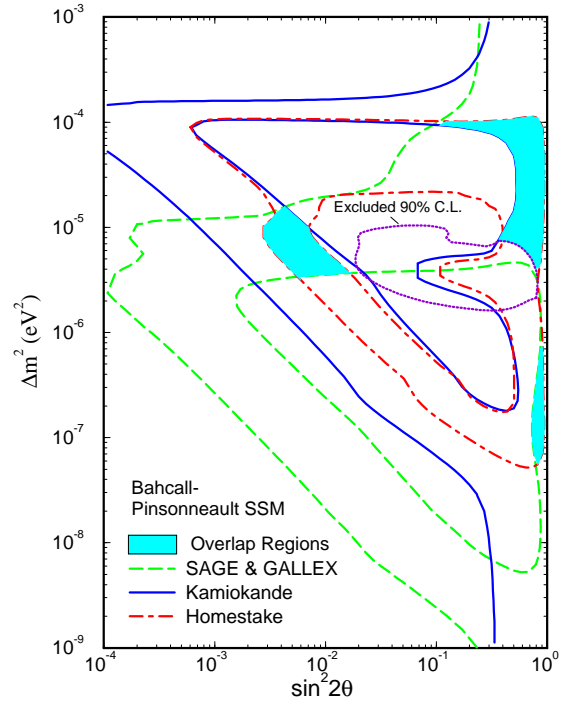

%
%\vspace*{4ex}                                                        %**epsf**
%\postscript{figures/p_comb_bpssm_overlap.ps}{\mswsize}               %**epsf**
%
\caption{
The combined allowed regions are simply taken as overlaps of the three
experimental constraints at 95\% C.L.  This C.L. definition overestimates
the uncertainties, allowing a parameter space which is marginally
allowed by different experiments, but, in fact, its $\chi^2$ is large
and the combined fit is poor.   As a result the obtained parameter space is
significantly overestimated compared to Fig.~\protect\ref{fig_MSW-earth},
especially in the large-angle region; even a third allowed region appears
in large-angle, small $\Delta m^2$.  Also this overlap procedure ignores
uncertainty correlations between different experiments.
}
\label{fig_MSW-overlap}
\end{figure}

%\clearpage                                                           %**epsf**
%
%                                  FIGURE
%
\begin{figure}[p]
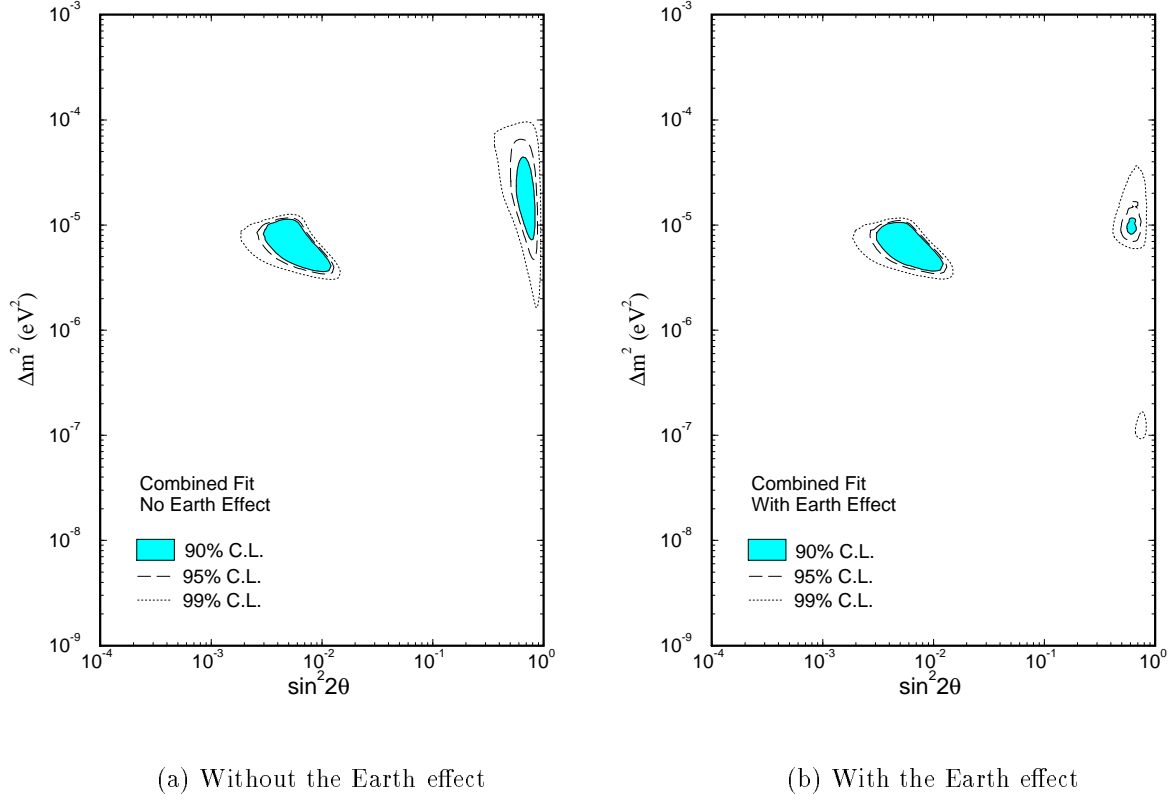

%
%\begin{center}                                                       %**epsf**
%\begin{tabular}{c}                                                   %**epsf**
%\setlength{\epsfxsize}{\mswsize}                                     %**epsf**
%\subfigure[Without the Earth effect]                                 %**epsf**
%{\epsfbox{figures/p_all_CLs_noearth.ps}}                             %**epsf**
%\subfigure[With the Earth effect]                                    %**epsf**
%      {\epsfbox{figures/p_all_CLs.ps}}                               %**epsf**
%\end{tabular}                                                        %**epsf**
%\end{center}                                                         %**epsf**
%
\caption{
The allowed regions of the combined experiments using an improved C.L
definition that assumes a gaussian probability density for each solution.
The 90, 95, and 99\% C.L. correspond to (a) $\Delta \chi^2$ = 5.9, 7.3,
and 10.6, ignoring the Earth effect, and (b) $\Delta \chi^2$ = 5.5, 7.0,
and 10.9, including the Earth effect.
}
\label{fig_MSW-newCL}
\end{figure}

% Here comes the figures ******************************************************
%\clearpage                                                          %**epsf**
%
%                                  FIGURE
%
%%\vspace*{-1ex}                                                     %**epsf**
\begin{figure}[p]
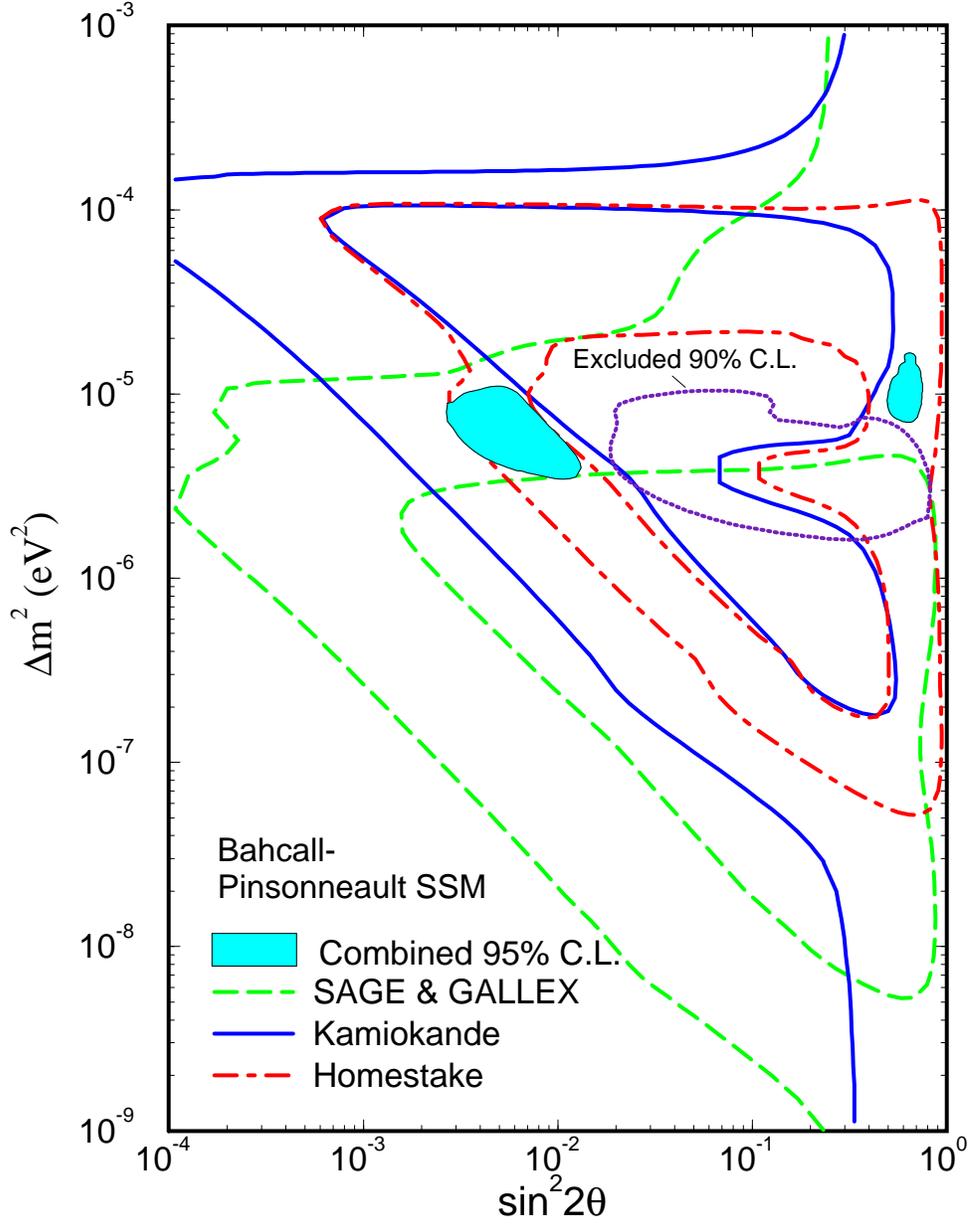

%\postscript{figures/p_comb_bpssm_best.ps}{0.85}                     %**epsf**
%
\caption{
The updated result of the combined MSW analysis assuming the
Bahcall-Pinsonneault SSM.  The Earth effect, Kamiokande II
day-night data, theoretical uncertainties and their
correlations are all included and the improved C.L. definition is
used.
}
\label{fig_MSW-best}
\end{figure}

%\clearpage                                                          %**epsf**
%
%                                  FIGURE
%
%%\vspace*{-1ex}                                                     %**epsf**
\begin{figure}[p]
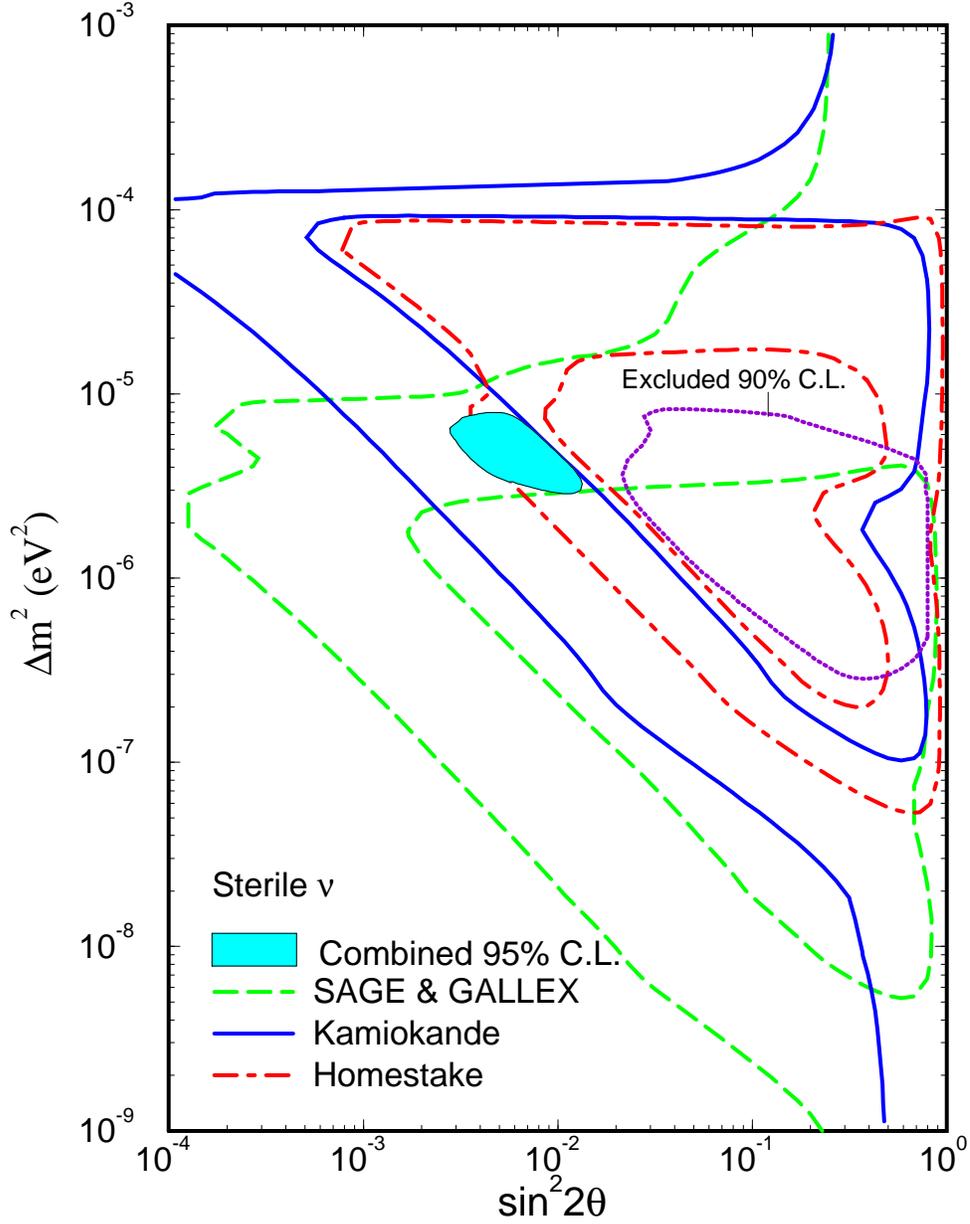

%\postscript{figures/p_comb_bpssm_best_sterile.ps}{0.85}             %**epsf**
%
\caption{
The updated result of the combined MSW analysis for the
oscillations into sterile neutrinos.  The Bahcall-Pinsonneault
SSM is assumed.
Bahcall-Pinsonneault SSM.  The Earth effect, Kamiokande II
day-night data, theoretical uncertainties and their
correlations are all included and the improved C.L. definition is
used.
}
\label{fig_MSW-best-sterile}
\end{figure}

%  The end of the figures *****************************************************

%**  Figures continue  ********************************************************

%\clearpage                                                           %**epsf**
%
%                                  FIGURE
%
\begin{figure}[p]
%
%\begin{center}                                                       %**epsf**
%\begin{tabular}{c}                                                   %**epsf**
%\setlength{\epsfxsize}{\mswsize}                                     %**epsf**
%\subfigure[Sterile $\nu$ without the Earth effect]                   %**epsf**
%                  {\epsfbox{figures/p_comb_sterile_noearth.ps}}      %**epsf**
%\subfigure[Sterile $\nu$ with the Earth effect]                      %**epsf**
%                  {\epsfbox{figures/p_comb_sterile.ps}}              %**epsf**
%\end{tabular}                                                        %**epsf**
%\end{center}                                                         %**epsf**
%
\caption{
The combined allowed regions for oscillations into sterile neutrinos
(a) without and (b) with the Earth effect.  No solutions are allowed in
the large-angle regions even at 99\% C.L.
}
\label{fig_sterile}
\end{figure}

%\clearpage                                                           %**epsf**
%
%                                  FIGURE
%
\begin{figure}[p]
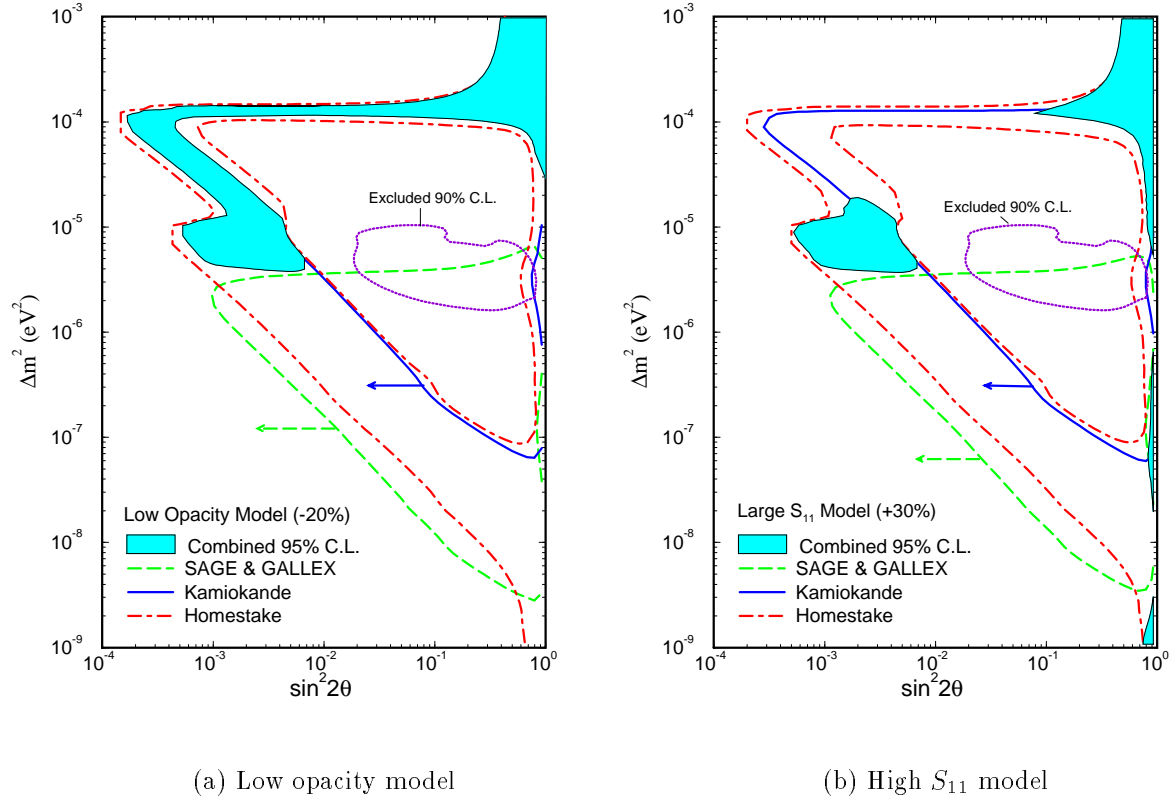

%
%\begin{center}                                                       %**epsf**
%\begin{tabular}{c}                                                   %**epsf**
%\setlength{\epsfxsize}{\mswsize}                                     %**epsf**
%\subfigure[Low opacity model]                                        %**epsf**
%{\epsfbox{figures/p_comb_lowop.ps}}                                  %**epsf**
%\subfigure[High $S_{11}$ model]                                      %**epsf**
%      {\epsfbox{figures/p_comb_highs11.ps}}                          %**epsf**
%\end{tabular}                                                        %**epsf**
%\end{center}                                                         %**epsf**
%
\caption{
The experimental constraints assuming nonstandard solar models that
predict neutrino fluxes {\em smaller} than the SSM:
(a) the opacity is reduced by 20\% at temperature larger than
$5 \times 10^6$ K \protect\cite{Dearborn}, (b) $S_{11}$ is increased by 30\%
\protect\cite{Castellani-etal}.
}
\label{fig_MSW-nonssms}
\end{figure}

%\clearpage                                                           %**epsf**
%
%                                  FIGURE
%
\begin{figure}[p]
%
%\begin{center}                                                       %**epsf**
%\begin{tabular}{c}                                                   %**epsf**
%\setlength{\epsfxsize}{\mswsize}                                     %**epsf**
%\subfigure[High Y Model]                                             %**epsf**
%{\epsfbox{figures/p_comb_highy.ps}}                                  %**epsf**
%\subfigure[Maximum rate model]                                       %**epsf**
%      {\epsfbox{figures/p_comb_maxrate.ps}}                          %**epsf**
%\end{tabular}                                                        %**epsf**
%\end{center}                                                         %**epsf**
%
\caption{
The experimental constraints assuming nonstandard solar models that
predict neutrino fluxes {\em larger} than the SSM:
(a) a high Y model \protect\cite{Bahcall-Ulrich} and (b) the maximum rate
model in which $S_{33}$ is artificially set to zero, maximizing the
gallium rate to 303 SNU \protect\cite{Bahcall-Pinsonneault}.
}
\label{fig_MSW-nonssms2}
\end{figure}

%%\clearpage                                                          %**epsf**
%
%                                  FIGURE
%
%\vspace*{4ex}                                                        %**epsf**
\begin{figure}[p]
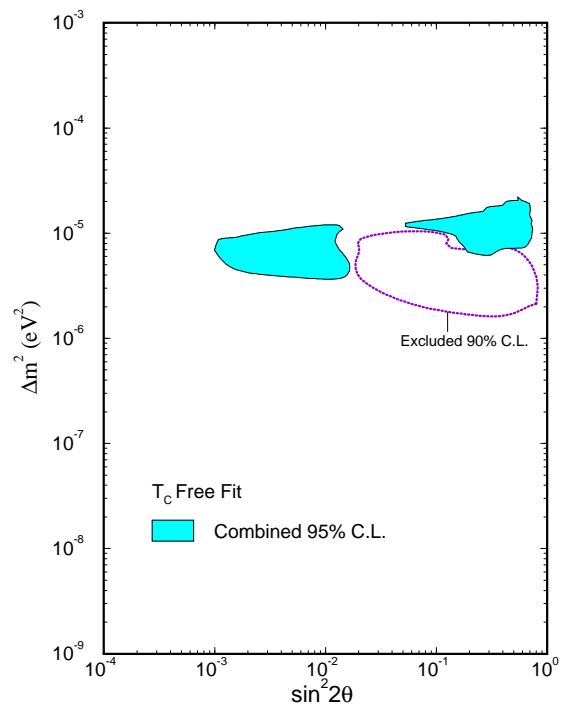

%\postscript{figures/p_comb_tcfree.ps}{\mswsize}                      %**epsf**
%
\caption{
The allowed regions of the combined observations when the central temperature
($T_C$) is a completely free parameter.  The combined data constrain
$T_C$ to $1.02 \pm 0.02$ ($1 \sigma$), which is consistent with the SSM
($T_C = 1 \pm 0.006$).  Also displayed is the region excluded by the
Kamiokande day-night data at 90\% C.L.; the exclusion comes from the
comparison between the different time bins, and is insensitive to the
absolute $^8$B flux or $T_C$.
}
\label{fig_MSW-Tcfree}
\end{figure}

%\newpage                                                             %**epsf**
%
%                                  FIGURE
%
\begin{figure}[p]
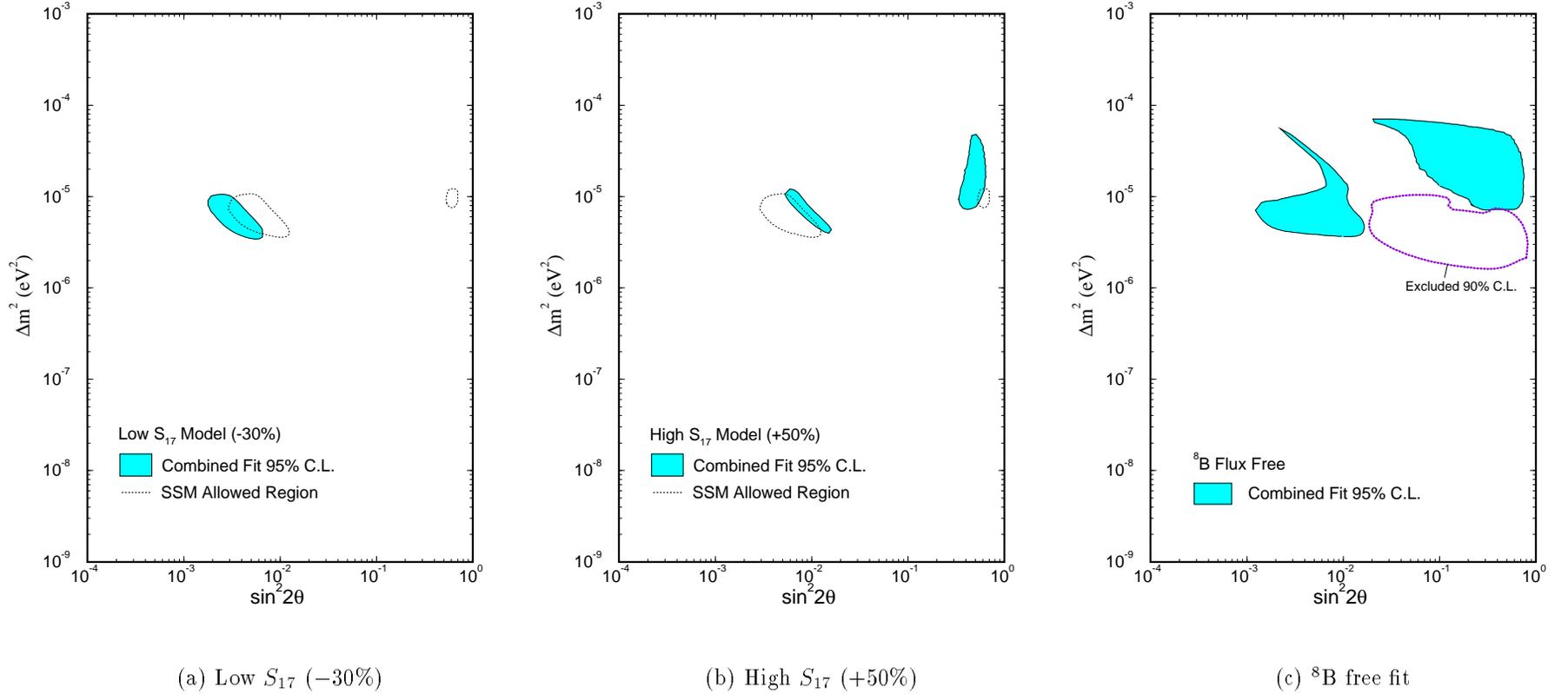

%
%\begin{center}                                                       %**epsf**
%\begin{tabular}{c}                                                   %**epsf**
%\setlength{\epsfxsize}{\mswsize}                                     %**epsf**
%\subfigure[Low $S_{17}$ ($-$30\%)]                                   %**epsf**
%{\epsfbox{figures/p_all_lows17.ps}}                                  %**epsf**
%\subfigure[High $S_{17}$ (+50\%)]                                    %**epsf**
%      {\epsfbox{figures/p_all_highs17.ps}}                           %**epsf**
%\subfigure[$^8$B free fit]                                           %**epsf**
%      {\epsfbox{figures/p_comb_Bfree.ps}}                            %**epsf**
%\end{tabular}                                                        %**epsf**
%\end{center}                                                         %**epsf**
%
\caption{
The experimental constraints when nonstandard $S_{17}$ values are assumed.
$S_{17}$ is directly proportional to the $^8$B flux and has the largest
uncertainty among the SSM input parameters.  We take $S_{17}$ to be (a)
30\% smaller than the SSM, (b) 50\% larger than the SSM.  In (c) the $^8$B
flux is treated as a completely free parameter; the combined data
constraint $^8\! \mbox{B} = 1.43 + 0.65 - 0.42$ of the standard value
($1 \sigma$) with $\chi^2_{\mbox{\protect\scriptsize min}}
= 2.0$ for 6 degrees of freedom.  Also shown in (c) is the excluded
region from the Kamiokande II day-night data (90\% C.L.), which is
independent of the $^8$B flux magnitude.
}
\label{fig_MSW-B8free}
\end{figure}

%\clearpage                                                           %**epsf**
%
%                                  FIGURE
%
\begin{figure}[p]
%
%\begin{center}                                                       %**epsf**
%\begin{tabular}{c}                                                   %**epsf**
%\setlength{\epsfxsize}{\mswsize}                                     %**epsf**
%\subfigure[Kam \& Cl combined]{\epsfbox{figures/p_comb_kam+cl.ps}}   %**epsf**
%\subfigure[Ga projection (1994)]{\epsfbox{figures/p_comb_94.ps}}     %**epsf**
%\end{tabular}                                                        %**epsf**
%\end{center}                                                         %**epsf**
%
\caption{
(a) The combined result of the Homestake and Kamiokande experiments.  From the
allowed regions we can predict  the gallium rate to be $\leq$ 100 SNU at
95\% C.L., which is consistent with the current observations of SAGE
and GALLEX.  (b) The combined result when the gallium experimental
uncertainty is reduced by a factor $1/\protect\sqrt{2}$, which is
equivalent to the data set through the end of 1994.  The central value of
the gallium rate is assumed to stay at the current value.  The present
values are used for the other experiments.
}
\label{fig_Ga-prediction}
\end{figure}

%\clearpage                                                           %**epsf**
%
%                                  FIGURE
%
\begin{figure}[p]
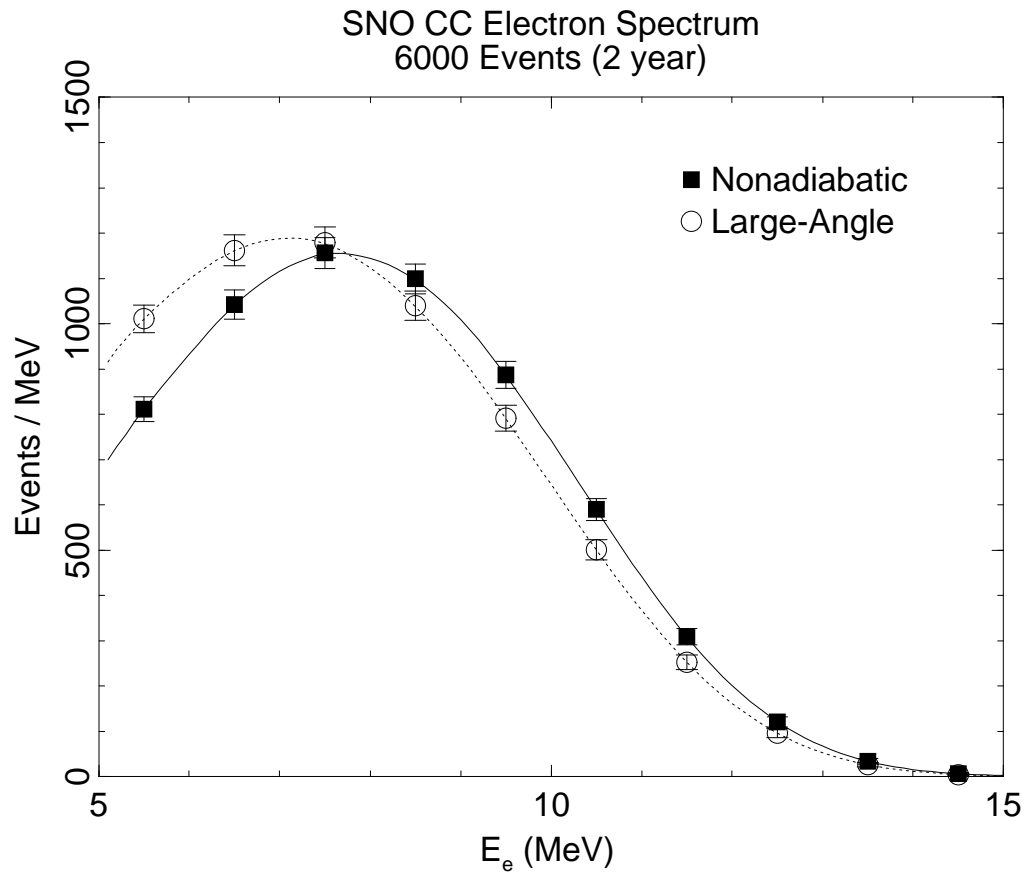

%
%\vspace*{-20ex}
%%%**epsf**
%\postscript{figures/sd_snocc.ps}{\specsize}
%%%**epsf**
%\vspace*{-15ex}
%%%**epsf**
%
\caption{
The SNO charged current spectrum expected for the nonadiabatic and
large-angle solution.  Astrophysical solutions would be similar to the
large-angle spectrum.  To compare the difference, the large-angle spectrum
is normalized to the the nonadiabatic.  The charged current cross section
\protect\cite{Nozawa,nu-d-cross-section} and the detector resolution
\protect\cite{SNO} are included.  The error bars indicate the statistical
uncertainties from 6,000 events (equivalent to two years of operation).
}
\label{fig_SNO-CCspectrum}
\end{figure}

%\clearpage                                                           %**epsf**
%                                  FIGURE
%
\begin{figure}[p]
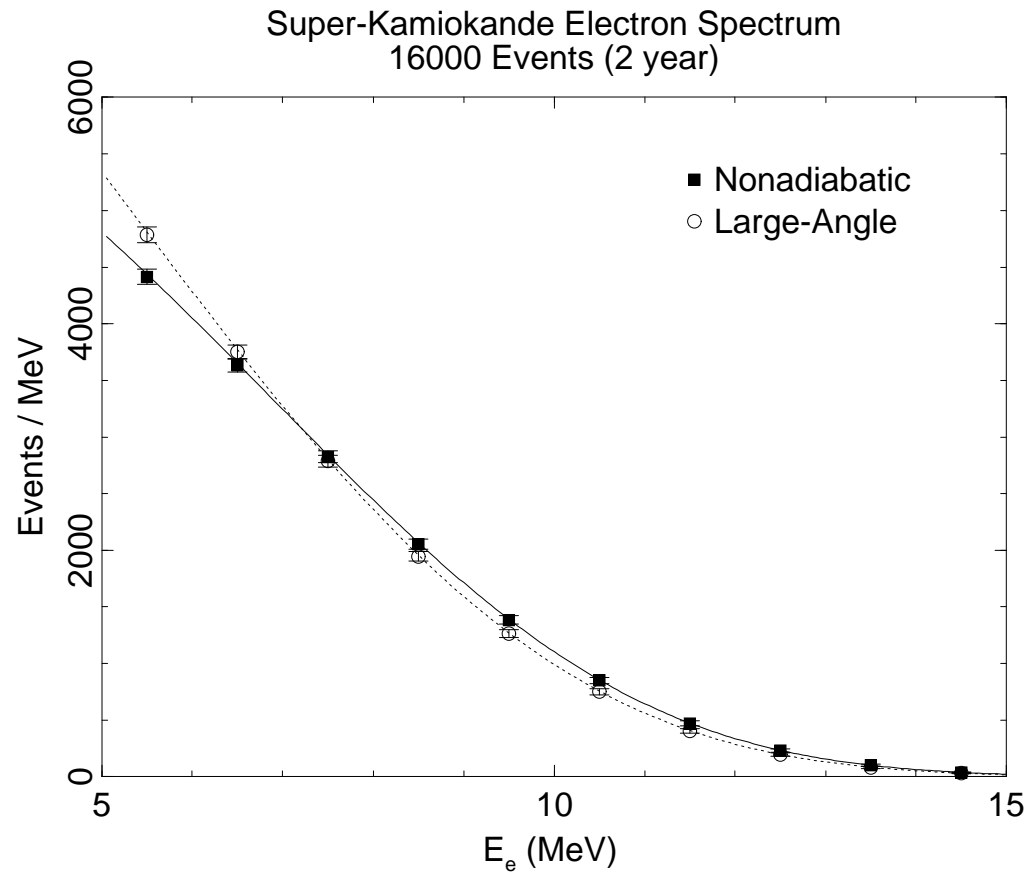

%
%\vspace*{-20ex}                                                      %**epsf**
%\postscript{figures/sd_superk.ps}{\specsize}                         %**epsf**
%\vspace{-15ex}                                                       %**epsf**
%
\caption{
The Super-Kamiokande electron spectrum expected for the nonadiabatic
and large-angle solutions.  To compare the difference, the large-angle
spectrum is normalized to the the nonadiabatic.  The detector resolution is
included \protect\cite{Super-Kamiokande}.  The error bars indicate the
statistical uncertainties from 16,000
events (equivalent to a two year operation).
}
\label{fig_SuperK-spectrum}
\end{figure}

%\clearpage                                                           %**epsf**
%
%                                  FIGURE
%
\begin{figure}[p]
%
%\vspace*{5ex}                                                        %**epsf**
%\begin{center}                                                       %**epsf**
%\begin{tabular}{c}                                                   %**epsf**
%\setlength{\epsfxsize}{\dnsize}                                      %**epsf**
%\subfigure[]{\epsfbox{figures/dn_snocc_na.ps}}                       %**epsf**
%\subfigure[]{\epsfbox{figures/dn_snocc_la.ps}}                       %**epsf**
%\end{tabular}                                                        %**epsf**
%\end{center}                                                         %**epsf**
%
\caption{
The Earth effect expected in SNO for the (a) nonadiabatic and (b)
large angle solution.  The night rate is shown with five bins according
to the angle between the direction to the Sun and the nadir at the detector.
The error bars indicate the statistical uncertainties equivalent to one year
of operation (3,000 events).
}
\label{fig_SNO-day-night}
\end{figure}

%\clearpage                                                           %**epsf**
%                                  FIGURE
%
\begin{figure}[p]
%
%\vspace*{5ex}                                                        %**epsf**
%\begin{center}                                                       %**epsf**
%\begin{tabular}{c}                                                   %**epsf**
%\setlength{\epsfxsize}{\dnsize}                                      %**epsf**
%\subfigure[]{\epsfbox{figures/dn_superk_na.ps}}                      %**epsf**
%\subfigure[]{\epsfbox{figures/dn_superk_la.ps}}                      %**epsf**
%\end{tabular}                                                        %**epsf**
%\end{center}                                                         %**epsf**
%
\caption{
The Earth effect expected in Super-Kamiokande for the (a) nonadiabatic
solution and (b) large angle solution.  The night rate is shown
with five bins according to the angle between the direction to the Sun and the
nadir at the detector.  The error bars indicate the statistical uncertainties
equivalent to one year of operation (8,000 events).
}
\label{fig_SuperK-day-night}
\end{figure}

%\clearpage                                                           %**epsf**
%
%                                  FIGURE
%
\begin{figure}[p]
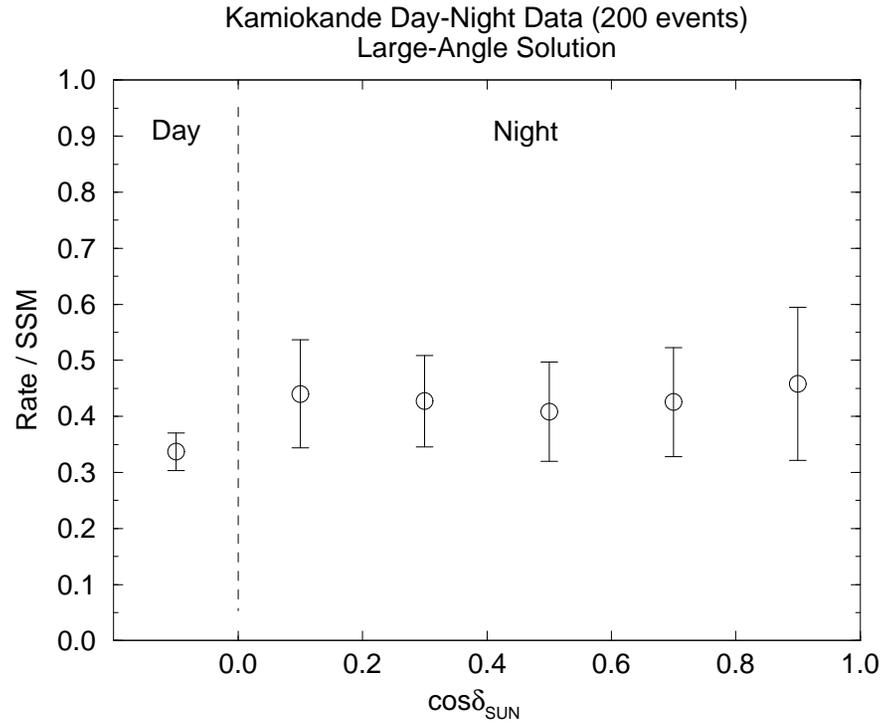

%
%\vspace*{10ex}                                                       %**epsf**
%\postscript{figures/dn_kam200_la.ps}{\dnsize}                        %**epsf**
%
%\vspace*{10ex}                                                       %**epsf**
\caption{
The Earth effect expected in Kamiokande for the (a) nonadiabatic
and (b) large angle solution.  The night time rates are shown
with five bins according to the angle between the direction to the Sun and the
nadir at the detector.  The error bars indicate the statistical uncertainties
assuming the total signals of 200 events.
}
\label{fig_KIII-day-night}
\end{figure}

%\clearpage                                                           %**epsf**
%
%                                  FIGURE
%
\begin{figure}[p]
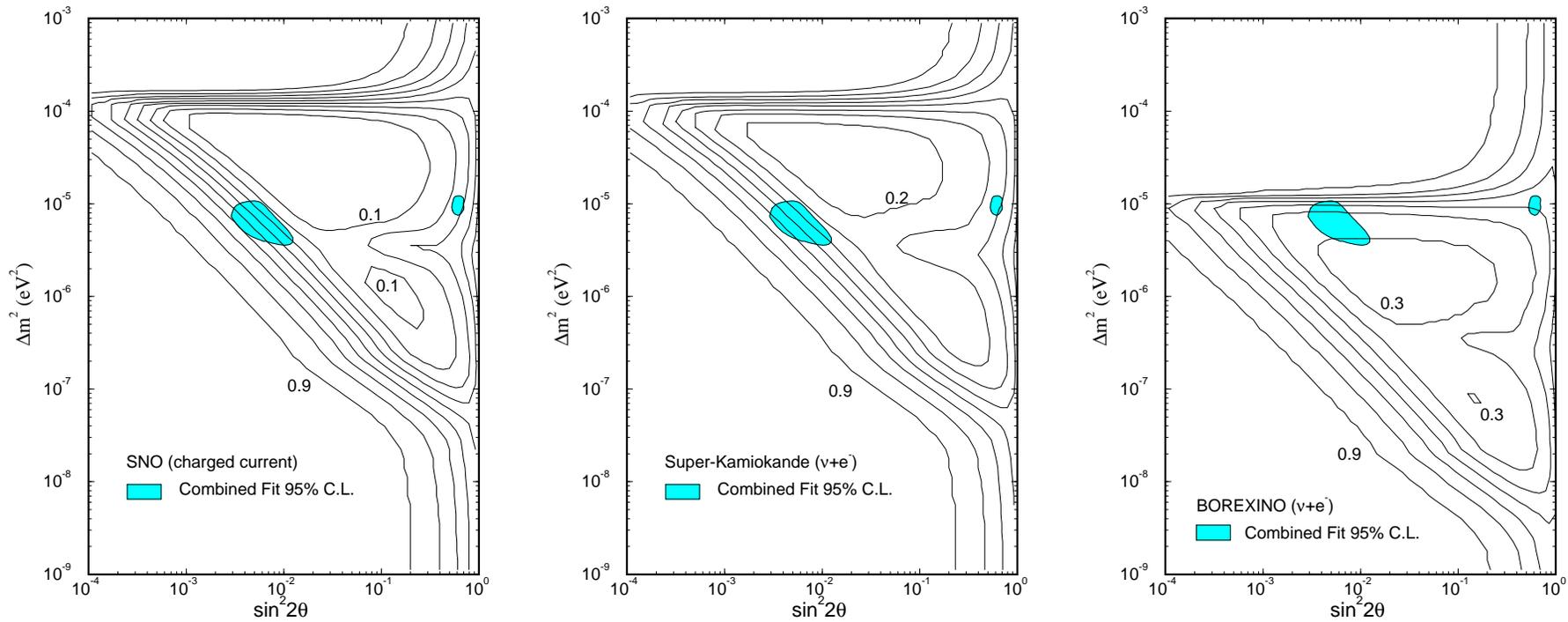

%
%\begin{center}                                                       %**epsf**
%\begin{tabular}{c}                                                   %**epsf**
%\setlength{\epsfxsize}{\mswsize}                                     %**epsf**
%\subfigure[SNO (charged current)]{\epsfbox{figures/c_snocc_comb.ps}} %**epsf**
%\subfigure[Super-Kamiokande]{\epsfbox{figures/c_superk_comb.ps}}     %**epsf**
%\subfigure[BOREXINO]{\epsfbox{figures/c_borexino_comb.ps}}           %**epsf**
%\end{tabular}                                                        %**epsf**
%\end{center}                                                         %**epsf**
%
\caption{
The contours of the signal to the SSM ratio including the time-averaged
Earth effect for (a) the SNO charged current
mode ($ \nu_e + d \rightarrow p + p + e $), (b) Super-Kamiokande ($\nu e$
scattering), and (c) BOREXINO ($\nu e$ scattering).  Superimposed are
the combined current allowed regions at 95\% C.L of the Homestake, Kamiokande,
and gallium experiments.
}
\label{fig_future-detectors}
\end{figure}

%  The end of the figures *****************************************************
% Here comes the figures ******************************************************
%
%                                  FIGURE
%
%\vspace*{-1ex}
\begin{figure}[p]
%\postscript{figures/big_picture_bw.ps}{0.80}                         %**epsf**
%\vspace{2ex}                                                        %**epsf**
%
\caption{
Shown are the constraints on the $\nu_e \leftrightarrow \nu_\mu$ and
$\nu_\mu \leftrightarrow \nu_\tau $ oscillations
\protect\cite{oscillation-limit}, the constraints on the oscillations into
sterile neutrinos ($\nu_e \leftrightarrow \nu_s$ and $\nu_\mu
\leftrightarrow \nu_s$) from
big-bang nucleosynthesis (BBN) \protect\cite{Shi-Schramm-Fields},
and three observational hints of neutrino mass: the MSW hypothesis
for the solar neutrino deficit, the oscillation interpretation of
the atmospheric neutrino deficit \protect\cite{atmospheric-neutrino-problem},
and the cold plus hot dark matter scenario for the COBE and large
scale structure data \protect\cite{CPHDM}.  Also displayed are predictions
from various theoretical models \protect\cite{BKL,%
Langacker-Neutrino-Telescopes,Cvetic-Langacker}.
}
\label{fig_big-picture}
\end{figure}

%  The end of the figures *****************************************************


\begin{thebibliography}{99}


\bibitem{Homestake}
R.~Davis, Jr., {\it et al.},
in {\it Proceedings of the 21th International Cosmic Ray Conference}, Vol. 12,
edited by R.~J.~Protheroe (University of Adelaide Press, Adelaide, 1990),
p.~143;
R.\ Davis, Jr.,
in {\it Frontiers of Neutrino Astrophysics},
edited by  Y.\ Suzuki and K.\ Nakamura (Universal Academy Press, Tokyo, 1993).

\bibitem{Homestake-update}
K.\ Lande, private communications; B.\ Cleveland, private communications.

\bibitem{Kamiokande}
K.~S.~Hirata {\it et al.}, Phys.\ Rev.\ Lett.\ {\bf 65}, 1297 (1990);
{\bf 65}, 1301(1990); {\bf 66}, 9 (1991);
%K.\ S.\ Hirata {\it et al.},
Phys.\ Rev.\ D {\bf 44}, 2241 (1991).

\bibitem{Kamiokande-update}
A.\ Suzuki,
KEK report No. 93-96, 1993 (unpublished).
%
%Y.\ Suzuki,
%in {\it International Symposium on Neutrino Astrophysics},
%Takayama, Kamioka, 1992 (unpublished).
Y.\ Suzuki,
in {\it Frontiers of Neutrino Astrophysics},
edited by  Y.\ Suzuki and K.\ Nakamura (Universal Academy Press, Tokyo, 1993)
p. 61.

\bibitem{SAGE}
A.~I.~Abazov, {\it et al.}, Phys.\ Rev.\ Lett.\ {\bf 67}, 3332 (1991).

\bibitem{SAGE-update}
%V.~N.~Gavrin, in {\em XXVI International Conference on High Energy
%Physics}, Dallas, 1992, edited by J.\ Sanford (AIP, New York, 1993).
%
V.\ N.\ Gavrin,
in {\em TAUP}, Gran Sasso, Italy, September, 1993.

\bibitem{GALLEX}
GALLEX Collaboration, P.~Anselmann {\it et al.},
Phys.\ Lett.\ B {\bf 285}, 376 (1992); {\bf 285}, 390 (1992).
% GALLEX II
P.\ Anselmann {\it et al.},
GALLEX collaboration Report No. GX 27a--1993
(to be published in  Phys.\ Lett.\ B).

\bibitem{Bahcall-Pinsonneault}
J.~N.~Bahcall and M.~H.~Pinsonneault,
Rev.\ Mod.\ Phys.\ {\bf 64}, 885 (1992).

\bibitem{Turck-Chieze-Lopes}
S.\ Turck-Chi\`eze and I.\ Lopes,
Astrophys. J. {\bf 408}, 347 (1993).
%
S.\ Turck-Chi\`{e}ze, S.\ Cahen, M.~Cass\'{e}, and C.\ Doom,
Astrophys.\ J.\ {\bf 335}, 415 (1988).

\bibitem{Bahcall-spectrum}
J.~N.~Bahcall,
Phys.\ Rev.\ D {\bf 44}, 1644 (1991).

\bibitem{BHL}
S.\ Bludman, N.\ Hata, and P.\ Langacker,
University of Pennsylvania preprint, UPR-0572T (1993).

\bibitem{BKL}
S.\ Bludman, D.\ Kennedy, and P.\ Langacker,
Phys.\ Rev.\ D {\bf 45}, 1810 (1992);
Nucl.\ Phys.\ B {\bf 374}, 373 (1992).

\bibitem{BHKL}
S.\ Bludman, N.\ Hata, D.\ Kennedy, and P.\ Langacker,
Phys.\ Rev.\ D {\bf 47}, 2220 (1993).

\bibitem{Bahcall-Bethe}
J.\ N.\ Bahcall and H.\ A.\ Bethe,
Phys.\ Rev.\ D {\bf 47}, 1298 (1993);
Phys.\ Rev.\ Lett.\ {\bf 65}, 2233 (1990);
H.\ A.\ Bethe and J.\ N.\ Bahcall,
Phys.\ Rev.\ D {\bf 44}, 2962 (1991).

\bibitem{MSW}
L.~Wolfenstein,
Phys.\ Rev.\ D {\bf 17}, 2369 (1978); {\bf 20}, 2634 (1979);

S.~P.~Mikheyev and A.~Yu.~Smirnov,
Yad.\ Fiz.\ {\bf 42}, 1441 (1985);
Nuo.\ Cim.\ {\bf 9C}, 17 (1986).

\bibitem{HL}
N.\ Hata and P.\ Langacker,
%University of Pennsylvania preprint UPR-0570T (1993).
Phys.\ Rev.\ D {\bf 48}, 2937 (1993).

\bibitem{Bahcall-Haxton}
J.\ N.\ Bahcall and W.\ C.\ Haxton,
Phys.\ Rev.\ D {\bf 40}, 931 (1989).

\bibitem{Shi-Schramm}
X.~Shi, D.~N.~Schramm, and J.~N.~Bahcall,
Phys.\ Rev.\ Lett.\ {\bf 69}, 717 (1992);
X.~Shi and D.~N.~Schramm,
Phys.\ Lett.\ B {\bf 283}, 305 (1992);
Fermilab preprint 92/322-A.

\bibitem{Gelb-Kwong-Rosen}
J.\ M.\ Gelb, W.\ Kwong, and S.\ P.\ Rosen,
Phys.\ Rev.\ Lett.\ {\bf 69}, 1864 (1992).

\bibitem{Krastev-Petcov}
P.~I.~Krastev and S.~T.~Petcov,
Phys.\ Lett.\ B {\bf 299}, 99 (1993).
%CERN preprint CERN-TH 6539/92.

\bibitem{Krauss-Gates-White}
L.~Krauss, E.~Gates, and M.~White,
Phys.\ Lett.\ B {\bf 299}, 94 (1993).
%Yale University preprint YCTP-P38-92.
%(The flux uncertainties are derived in
%M.\ White, L.\ Krauss, and E.\ Gates,
%Phys.\ Rev.\ Lett.\ {\bf 70}, 375 (1993).)
In this paper, the input data of the Kamiokande and Homestake, SAGE, and
GALLEX are respectively $ 0.49 \pm 0.08 $, $ 0.28 \pm 0.04 $,
$ 0.44 \pm 0.21 $, and $ 0.63 \pm 0.16 $, which is slightly
different from our updated input data (Table~\ref{tab_exps}).

\bibitem{Bahcall-Ulrich}
J.~N.~Bahcall and R.~N.~Ulrich,
Rev.\ Mod.\ Phys.\ {\bf 60}, 297 (1988);
J.~N.~Bahcall, {\it Neutrino Astrophysics}, (Cambridge University
Press, Cambridge, England, 1989).

\bibitem{Bahcall-Texas}
J.\ N.\ Bahcall,
Institute of Advanced Study preprint IASSNS 92/54 (1992)
(To be published in the {\it Proceedings of the XXVI International Conference
on High Energy Physics}, Dallas, Texas (1992).)

\bibitem{Earth-effect}
A.\ J.\ Baltz and J.\ Weneser,
Phys.\ Rev.\ D {\bf 35}, 528 (1987);  {\bf 37}, 3364, (1988);
E.\ D.\ Carlson,
Phys.\ Rev.\ D {\bf 34}, 1454 (1986);
J.\ Bouchez, M.\ Cribier, W.\ Hampel, J.\ Rich, M.\ Spiro, and D.\ Vignaud,
Z.\ Phys.\ C {\bf 32}, 499 (1986);
M.\ Cribier, W.\ Hampel, J.\ Rich, and D.\ Vignaud,
Phys.\ Lett.\ B {182}, 89 (1986);
S.\ Hiroi, H.\ Sakuma, T.\ Yanagida, and M.\ Yoshimura,
Prog. Theor. Phys. {\bf 78}, 1428 (1987);
A.\ Dar, A.\ Mann, Y.\ Melina, and D.\ Zajfman,
Phys.\ Rev.\ D {\bf 35} 3607, (1987);
P.\ I.\ Krastev and S.\ T. Petcov,
Phys.\ Lett.\ B {\bf 205}, 84 (1988);
R.\ S.\ Raghavan {\it et al.},
Phys.\ Rev.\ D {\bf 44} 3786, (1991);
J.\ M.\ LoSecco,
Phys. Rev. D {\bf 47}, 2032 (1993);
V.\ Barger, K.\ Whisnant, S.\ Pakvasa, and R.\ J.\ N.\ Phillips,
Phys.\ Rev.\ D {\bf 22}, 2718 (1980).

\bibitem{T_c-error}
See Figure 11 of Bahcall and Ulrich \cite{Bahcall-Ulrich} and  Figures 6.2
(p.\ 150) and 6.3 (p. 152) of Bahcall \cite{Bahcall-Ulrich}.

\bibitem{Bahcall-logderi}
See Table XV of Bahcall and Ulrich \cite{Bahcall-Ulrich} or Table 7.2 (p. 184)
of Bahcall \cite{Bahcall-Ulrich}.

\bibitem{Kennedy}
D.\ C.\ Kennedy,
University of Pennsylvania preprint UPR 0442-T(REV) (1992).

\bibitem{Parke}
S.\ J.\ Parke,
Phys.\ Rev. Lett. {\bf 57}, 1275 (1986).

\bibitem{Pizzochero}
P.\ Pizzochero,
Phys.\ Rev.\ D {\bf 36}, 2293 (1987).

\bibitem{Petcov}
S.\ T.\ Petcov,
Phys.\ Lett.\ B {\bf 200}, 373 (1988);
Nucl.\ Phys.\ B (Proc. Suppl.) 13, 527 (1990).

\bibitem{Haxton}
W.\ C.\ Haxton,
Phys.\ Rev.\ D {\bf 35}, 2352 (1987).

\bibitem{Kuo-Pantaleone}
T.\ K.\ Kuo and J.\ Pantaleone,
Rev.\ Mod.\ Phys.\ {\bf 61}, 937 (1989).

\bibitem{Kam-daynight}
K.~S.~Hirata {\it et al.}, Phys.\ Rev.\ Lett.\  {\bf 66}, 9 (1991).

\bibitem{sterile}
P.\ Langacker,
University of Pennsylvania Report No., UPR 0401T (1989);
R.\ Barbieri and A.\ Dolgov,
Nucl.\ Phys.\ B {\bf 349}, 743 (1991);
K.\ Enqvist, K.\ Kainulainen, and J.\ Maalampi,
Phys.\ Lett.\ B {\bf 249}, 531 (1990);
M.\ J.\ Thomson and B.\ H.\ J.\ McKellar,
Phys.\ Lett. B {\bf 259}, 113 (1991);
V.\ Barger {\it et al.},
Phys.\ Rev.\ D {\bf 43}, 1759 (1991);
P.\ Langacker and J.\ Liu,
Phys.\ Rev.\ D {\bf 46}, 4140 (1992).

\bibitem{Shi-Schramm-Fields}
X.\ Shi, D.\ Schramm, and B.\ Fields,
Phys.\ Rev.\ D {\bf 48}, 2563 (1993).

\bibitem{Dearborn}
D.\ Dearborn,
private communications.

\bibitem{Castellani-etal}
V.\ Castellani, S.\ Degl'Innocenti, and G.\ Fiorentini,
Phys.\ Lett.\ B {\bf 303}, 68 (1993).

\bibitem{Ga-prediction}
A similar prediction was given previously by
J.\ N.\ Bahcall and H.\ A.\ Bethe,
Phys.\ Rev.\ Lett.\ {\bf 65}, 2233 (1990);
A.\ J.\ Baltz and J.\ Weneser,
Phys.\ Rev.\ D {\bf 66}, 520 (1991).
We consider it worthwhile to update the prediction by using the
latest Homestake and Kamiokande results and including the theoretical
uncertainties.

\bibitem{SNO}
G.\ T.\ Ewan {\it et al.}
``Sudbury Neutrino Observatory Proposal'', Report No.\ SNO-87-12, 1987
(unpublished);
``Scientific and Technical Description of the Mark II SNO Detector'',
edited by E.\ W.\ Beier and D.\ Sinclair, Report No.\ SNO-89-15, 1989
(unpublished).

\bibitem{Super-Kamiokande}
Y.\ Totsuka,
University of Tokyo (ICRR) Report No.\ ICCR-Report-227-90-20, 1990
(unpublished).

\bibitem{BOREXINO}
``BOREXINO at Gran Sasso --- proposal for a real time detector for
low energy solar neutrinos'', Vol.\ 1,
edited by G.\ Bellini, M.\ Campanella, D.\ Giugni, and R.\ Raghavan (1991).

\bibitem{ICARUS}
C.\ Rubbia,
Report No.\ CERN-PPE/93-08, 1993 (unpublished).

\bibitem{Nozawa}
S.\ Nozawa,
private communications.

\bibitem{nu-d-cross-section}
M.\ Doi and K.\ Kubodera,
Phys. Rev.\ C {\bf 45}, 1988 (1992);
S.\ Ying, W.\ Haxton, and E.\ Henley
Phys. Rev.\ C {\bf 45}, 1982 (1992);
K.\ Kubodera and S.\ Nozawa,
University of South Carolina Report No.\ USC(NT)-93-6 (unpublished)
and references therein.

\bibitem{LoSecco}
J.\ M.\ LoSecco \cite{Earth-effect}.

\bibitem{atmospheric-neutrino-problem}
See, for example,
E.\ Beier {\it et al.},
Phys.\ Lett.\ B{283}, 446 (1992),
and references therein.

\bibitem{CPHDM}
G.\ Smoot {\it et al.},
Astrophys.\ J.\ {\bf 396}, L1 (1992);
%
E.\L.\ Wright {\it et al.},
Astrophys.\ J.\ {\bf 396}, L13 (1992);
%
R.\ Schaefer and Q.\ Shafi,
Bartol Research Institute Report No.\ BA-93-52;
Nature {\bf 359}, 199 (1992);
%
A.\ Klypin {\it et al.},
University of California (Santa Cruz) Report No.\  SCIPP-92-52;
%
R.\ L.\ Davis {\it et al.},
Phys.\ Rev.\ Lett.\ {\bf 69}, 1856 (1992).

\bibitem{Langacker-Neutrino-Telescopes}
P.\ Langacker,
University of Pennsylvania Report No.\ UPR-0511T, 1992 (unpublished).
%{\it Proceedings of the 4th International Symposium on Neutrino Telescopes},
%Venice, Italy, 1992,
%edited by M.\ Bald-Ceolin (unpublished).

\bibitem{Cvetic-Langacker}
M.\ Cveti\v{c} and P.\ Langacker,
Phys.\ Rev.\ D {\bf 46}, R2759 (1992).

\bibitem{oscillation-limit}
G.\ Bernardi,
in {\it XXIV International Conference on High Energy Physics} edited by
R.\ Kotthaus and J.\ K\"uhn (Springer, Berlin, 1989) p. 1076.

\end{thebibliography}
\end{document}